\documentclass[11pt]{article}
\usepackage[a4paper, total={5in, 9in}]{geometry}

\usepackage{amsmath,amssymb,dsfont,amsfonts,amsthm,amscd}
\usepackage{tabularx}
\usepackage{graphicx}
\usepackage{subcaption}
\usepackage{times}
\usepackage[table]{xcolor}
\usepackage{float}
\usepackage[pdfencoding=auto]{hyperref}
\usepackage{url}
\usepackage{array}
\usepackage{bm}
\usepackage{authblk}
\usepackage{natbib}
\usepackage{setspace}
\usepackage{booktabs}
\usepackage{rotating}
\usepackage{multirow}

\DeclareMathOperator{\vect}{vec}
\DeclareMathOperator{\Tr}{Tr}

\newtheorem{theorem}{Theorem}[section]
\newtheorem{remark}[theorem]{Remark}

\title{A conic optimization framework for shape-constrained functional regression analysis}

\author[1]{Mar\'ia Durb\'an}
\author[1]{Vanesa Guerrero}
\author[2]{Cristina Molero-R\'io \thanks{Corresponding author: mmolero@us.es}\,\,}
\affil[1]{Department of Statistics, Universidad Carlos III de Madrid, Madrid, Spain}
\affil[2]{Department of Statistics and Operations Research \& Instituto de Matem\'aticas de la Universidad de Sevilla, Universidad de Sevilla, Seville, Spain}

\date{}

\providecommand{\keywords}[1]{\textbf{\textit{Keywords---}} #1}

\begin{document}

\maketitle

\begin{abstract} 
Standard regression methods do not always guarantee that the estimated model reflects prior knowledge about how predictors relate to the response, which motivates constrained regression. When the effects to be estimated are smooth nonlinear functions, a natural family of constraints is that of shape constraints, such as sign, monotonicity, or curvature, which encode qualitative properties that these functions are often expected to satisfy. While shape-constrained estimation of smooth functions is well established in classical regression with scalar covariates and response, its extension to functional regression, where the coefficients are themselves functions, remains largely unexplored.
In this paper, we propose the functional shape-constrained conic P-spline estimator (FunSCOPE), a unified framework for shape-constrained functional linear regression that covers scalar-on-function, function-on-scalar, and function-on-function models, and in which any combination of the shapes above can be imposed on the functional regression coefficient. Building on a P-spline approach, the desired shapes are expressed as conditions on the scalar regression coefficients, which leads to an estimation problem formulated as a conic optimization problem. For univariate coefficients, as in scalar-on-function, function-on-scalar, and a particular case of the function-on-function regression, the concurrent case, these conditions are both necessary and sufficient, in contrast to existing approaches, which provide only sufficient conditions; for bivariate coefficients, as in general function-on-function models, they are neither necessary nor sufficient. Numerical experiments show that FunSCOPE outperforms competing methods in terms of estimation error and computational time.
\end{abstract}


\section{Introduction}
\label{cap:Introduction}

The increasing availability of data collected continuously over time or space has driven growing interest in functional data analysis (FDA), a branch of statistics that treats observations as realizations of functions over a continuum rather than as finite-dimensional vectors \citep{aneiros2022functional, gertheiss2024functional, wang2016functional}. 
FDA methods have found applications across diverse fields, including biomedicine and  health sciences \citep{leng2006functional,strzalkowska2021censored}, chemistry \citep{blanquero2016extreme},  meteorology \citep{james2009functional}, and manufacturing and service operations \citep{park2022motion}.

Since the pioneering work of \citet{ramsay1997functional}, FDA has expanded considerably to address inferential and predictive tasks, extending classical multivariate techniques, such as principal component analysis, clustering, and classification, to the functional setting.  In this work, our focus is on functional linear regression \citep{morris2015functional}.  Depending on whether the functional nature resides in the predictors, the response, or both, one distinguishes between scalar-on-function \citep{reiss2017methods}, function-on-scalar, and function-on-function regression \citep{greven2017general}, respectively.

Functional linear regression involves the estimation of functional regression coefficients, which live in infinite-dimensional spaces and must therefore be approximated in practice. Two main approaches have been proposed to this end: nonparametric methods, such as kernel smoothing, which estimate the functional object pointwise by locally weighting observations \citep{ferraty2006nonparametric}; and basis expansion methods, which represent the functional object as a linear combination of basis functions, such as Fourier bases, wavelets, Bernstein polynomials, or splines, reducing the estimation problem to that of determining a finite set of coefficients \citep{ramsay1997functional}. Among the latter, we use a reduced-rank basis of B-spline functions to represent the functional objects involved in functional regression models, namely functional covariates, responses, and regression coefficients, and fit them via a penalized spline (P-spline) approach \citep{eilers1996flexible, 20years}. P-splines have emerged as a particularly attractive choice in FDA because they combine a flexible basis with a roughness penalty that controls smoothness independently of the number of basis functions or knots used, avoiding careful knot selection while remaining computationally tractable. 

Flexibility alone, however, is not always sufficient: purely data-driven estimation does not ensure that the fitted model is consistent with what is known about the problem. In high-stakes domains, predictive models are not just analytical tools, but inform decisions with real consequences \citep{gdpr2016, euaiact2024, rudin2019}. In these contexts, reliability means more than low prediction error: it means models that behave sensibly, respect known physical or domain constraints, and do not produce implausible outputs. For instance, a predicted mortality rate should not be negative, a drug efficacy curve should reflect the known monotone improvement of a patient’s condition over time, and a production function should not violate concavity. Purely data-driven models, however, do not inherently guarantee such properties, and may yield untrustworthy outputs, particularly when data are scarce, noisy, or unrepresentative \citep{rudin2022}. This motivates constrained regression, where structural knowledge is embedded into the estimation process to ensure that models are not only accurate but also reliable.

Mathematical optimization has proven to be a powerful and flexible tool for constrained regression, with successful applications across a wide range of settings, including sparsity \citep{carrizosa2017sparsity, navarro2025feature}, fairness \citep{blanquero2022sparse}, and cost-sensitivity \citep{blanquero2021costsensitive, carrizosa2026prescriptive}. When the effects to be estimated are smooth nonlinear functions, a particularly relevant family of constraints is that of shape constraints, which encode qualitative properties of these functions, such as their sign, monotonicity, or curvature. Shape-constrained estimation of smooth functions has been studied, and proven effective, in classical regression with scalar covariates and response \citep{curmei2025, espana2024, lin2022, meyer2018, Navarro2023AMC, navarro2024, pya2015}. However, its extension to functional regression, where the coefficients are themselves functions, remains largely unexplored. The few existing contributions include \citet{ghosal2023shape}, who propose shape-constrained estimation for scalar-on-function, function-on-scalar, and function-on-function regression using Bernstein polynomials; and \citet{han2025inference}, who develop inference procedures for partially shape-constrained function-on-scalar regression using kernel and spline methods, under local, subinterval-specific shape constraints, in contrast to the global, whole-domain constraints considered here and in \citet{ghosal2023shape}.

In this paper, we propose the functional shape-constrained conic P-spline estimator (FunSCOPE), a unified framework for shape-constrained functional linear regression that encompasses scalar-on-function, function-on-scalar, and function-on-function models, and in which any combination of these shapes can be imposed on the functional regression coefficient. A key observation underlying our approach is that all these regression settings can be expressed within a common P-spline framework. This establishes a direct connection with the conic optimization approaches of \citet{Navarro2023AMC, navarro2024}, which we extend to the functional regression setting: since in the B-spline representation the shape of the fitted function is directly linked to its coefficients, the desired shapes can be expressed as conditions on those coefficients, and the resulting estimation problem is a conic optimization problem.

For univariate functional regression coefficients, as in scalar-on-function, function-on-scalar, and a particular case of the function-on-function regression, the concurrent case, these conditions are both necessary and sufficient for the desired shapes to hold. For bivariate coefficients, as in general function-on-function regression, they are neither necessary nor sufficient; nevertheless, the smoothness imposed by the P-spline penalty prevents abrupt changes and makes violations of the shape constraints unlikely in practice. In both the univariate and bivariate cases, this contrasts with \citet{ghosal2023shape}, who formulate shape constraints as linear inequality constraints solved via quadratic programming, and \citet{han2025inference}, who rely on cone projection for spline-based estimation and on discretized grid constraints for kernel-based estimation. In both cases, the resulting constraints are sufficient but not necessary, which restricts the estimator to a subset of the truly shape-satisfying functions. Moreover, unlike \citet{han2025inference}, who focus exclusively on function-on-scalar regression and require specific spline degrees depending on the type of constraint, our framework provides a unified treatment of all three functional regression settings and is valid for splines of any degree. Although our framework is formulated for global shape constraints, that is, imposed over the whole domain of the functional data, it also accommodates the local setting of \citet{han2025inference} as a special case: restricting the constraints to selected subintervals simply reduces the number of constraints in the optimization problem.

Our approach also differs from these works in how the data are represented. We assume a basis representation of the functional data themselves, treating functional observations as genuinely functional objects rather than as high-dimensional discrete vectors \citep{ghosal2023shape, han2025inference}, in line with the standard treatment of functional data in FDA \citep{ramsay1997functional}. Once each curve has been represented in the basis, estimation operates on the resulting basis coefficients instead of the raw discrete measurements. Consequently, our procedure does not require all individuals to be observed at the same time points: each curve may be recorded at its own set of instants, possibly differing in number and location, as long as all curves share a common domain. Approaches that work directly with values on a grid require a common set of observation points, or a preliminary smoothing step to recover the curves on a common grid \citep{ghosal2023shape}; in our framework, this is handled directly through the basis representation of each curve, which makes it applicable to irregularly or asynchronously sampled functional data, a situation frequently encountered in practice \citep{yao2005functional}. Treating the data as functions also has technical consequences, for instance in the concurrent function-on-function model, which are addressed in Section~\ref{subsubsec:concurrent}.

The remainder of the paper is organized as follows. Section \ref{sec:FRPsplines} introduces P-splines for functional regression models. Section \ref{sec:Shape-FR} presents FunSCOPE, including the conditions on the spline coefficients that enforce the desired shapes and the resulting conic optimization formulation. Section \ref{Computational experiments} reports the computational experiments, in which FunSCOPE is compared with competing methods in terms of estimation error and computational time. Finally, Section \ref{Conclusions and perspectives} presents the conclusions and possible lines of future research.

\section{Functional regression with P-splines \label{sec:FRPsplines}}

In this section, we first review penalized splines (P-splines) \citep{eilers1996flexible, 20years} for estimating smooth regression models in which both the covariate and the response are scalar (Section \ref{subsec:psplines}). We then show how P-splines extend to functional linear regression, where the covariate, the response, or both are functions, in Sections \ref{subsec:SOF}, \ref{subsec:FOS}, and~\ref{subsec:FOF}, respectively. Throughout, functional observations are treated as functional objects, rather than as high-dimensional discrete vectors: they are represented through a basis expansion and smoothed via a roughness penalty on the basis coefficients, following the P-spline approach.

Despite their different natures, all these regression settings can be expressed within a common P-spline framework, in which the regression coefficients are estimated by solving a penalized least-squares problem of the form
\begin{equation}\label{eq:frm}
    \min_{\boldsymbol{\delta} } \quad  \| \mathbf{u} - \mathbf{Z} \boldsymbol{\delta} \|^2_2 + P(\boldsymbol{\delta}),\\
\end{equation}
where $\|\cdot\|_2$ refers to the Euclidean norm, and the explicit forms and dimensions of the response vector $\mathbf{u}$, the design matrix $\mathbf{Z}$, the regression coefficients $\boldsymbol{\delta}$, and the penalty term $P(\boldsymbol{\delta})$ depend on the specific regression model, as described below.

\subsection{Smooth regression with P-splines \label{subsec:psplines}}
Let $\{(x_i,y_i)\}_{1\leq i\leq n}$ be a set of observations from a scalar covariate $X$ and a scalar response variable $Y.$ Assuming $x_1\leq x_2\leq \ldots \leq x_n$, the goal in smooth regression is to estimate a curve $f: \left[x_1, x_n\right] \longrightarrow \mathbb{R}$ such that
\begin{equation}\label{eq:sosr}
    y_i = f(x_i) + \varepsilon_i,
\end{equation}
where the error terms $\varepsilon_i$ are independent and identically distributed samples from a Gaussian distribution, $i=1,\ldots,n.$ 

Estimating $f$ in \eqref{eq:sosr} using a P-spline approach involves representing $f$ using B-spline basis functions \citep{DeBoor1978ASplines} and then solving a penalized least-squares problem. Specifically, $f$ in \eqref{eq:sosr} is represented as
\begin{equation}\label{eq:sosrBspline}
    f(x) = \displaystyle\sum_{j=1}^{k+d} \theta_j B_j(x) = \mathbf{B}(x) \boldsymbol{\theta},
\end{equation}
where $B_j(\cdot),$ are B-spline basis functions of degree $d$ defined over a non-decreasing sequence of real numbers (the knots)  $\{\tau_\ell\}_{1\leq\ell\leq k+2d+1},$  such that $\tau_{d+1}=x_1,$  $\tau_{k+d+1} =x_n,$ and $k$ is the number of intervals in which the domain $[x_1,x_n]$ is split; and $\theta_j$ are the regression coefficients, $j=1,\ldots,k+d.$ In other words, $\mathbf{B}(\cdot) =(B_1(\cdot),\ldots,B_{k+d}(\cdot))$ is the vector of B-spline basis functions and $\boldsymbol{\theta}= (\theta_1,\ldots,\theta_{k+d})^\top$ is the vector of regression coefficients. Because B-spline basis functions are piecewise polynomials of degree $d$, so is $f$ in \eqref{eq:sosrBspline}.  

The P-spline approach estimates the regression coefficients in \eqref{eq:sosrBspline} by solving the following penalized least-squares problem:
\begin{equation}\label{eq:sosrPspline}
    \displaystyle \min_{\boldsymbol{\theta} \in \mathbb{R}^{k+d}} \, \sum_{i=1}^n \left(y_i - \mathbf{B}(x_i) \boldsymbol{\theta} \right)^2 + P(\boldsymbol{\theta}),
\end{equation}
where the first term corresponds to the sum of squared errors and the second, $P(\cdot)$, is a penalty term that controls the smoothness of the curve $f$ in \eqref{eq:sosrBspline}. Several choices are available for this penalty term, a common option being one based on finite differences between adjacent coefficients. Adopting this penalty turns \eqref{eq:sosrPspline} into an unconstrained convex quadratic problem, expressed in matrix form as:
\begin{equation}\label{eq:sosrPspline-matrix}
    \displaystyle \min_{\boldsymbol{\theta} \in \mathbb{R}^{k+d}} \,  \left\|\mathbf{y} -  \mathbf{B}\boldsymbol{\theta}\right\|_2^2 + \lambda \left\|\mathbf{D}\boldsymbol{\theta} \right\|_2^2,
\end{equation}
where $\mathbf{y}=(y_1,\ldots,y_n)^\top$ is the vector of response observations; $\mathbf{B}$ the $n\times (k+d)$ design matrix of the regression model with $(i,j)-$th entry $B_j(x_i)$; $\mathbf{D}$ the second-order difference matrix of size $(k+d-2)\times (k+d);$ and $\lambda \geq 0$  a smoothing parameter that controls the smoothness of the fitted curve. 
Therefore, the smooth regression model \eqref{eq:sosr} estimated with P-splines is expressed as in \eqref{eq:frm}, where $\mathbf{u} = \mathbf{y}\in \mathbb{R}^n,$ $\mathbf{Z} = \mathbf{B}\in\mathbb{R}^{n\times(k+d)},$ $\boldsymbol{\delta} = \boldsymbol{\theta}\in\mathbb{R}^{k+d},$ and $P(\boldsymbol{\delta}) = \lambda\|\mathbf{D}\boldsymbol{\delta}\|_2^2.$ 

\subsection{Scalar-on-function regression with P-splines \label{subsec:SOF}}

Let $\{(x_i(t), y_i)\}_{1\leq i\leq n}$ be a set of observations from a functional covariate $X(t)$, $t\in[0,T]\subset \mathbb{R}^+$, and a scalar response variable $Y.$ Then, the scalar-on-function (SOF) linear regression model \citep[see, e.g.,][]{cardot1999functional, ramsay1997functional} is given by 
\begin{equation}\label{eq:sof}
    y_i = \beta_0 + \int_0^T x_i(t)\beta(t)\,dt + \varepsilon_i,
\end{equation}
where the error terms $\varepsilon_i$ are independent and identically distributed samples from a Gaussian distribution, $i=1,\ldots,n$. Assuming that the sample data is mean-centered at zero, $\beta_0 =0,$ and $\beta(t),$  $t\in[0,T],$  is the functional regression coefficient to be estimated.

In practice, the functional covariate is observed at a discrete set of points. For simplicity of exposition, we assume that all individuals are observed at the same instants $t_m \in [0,T]$, $m=1,\ldots,M$. This is not required: the observation instants, and their number, may differ across individuals, so that the sample for the $i$-th curve is $\{(t_{im}, x_i(t_{im}))\}_{1\leq m\leq M_i}$, as long as all curves share the common domain $[0,T]$. Despite the discrete nature of the observations, we treat $x_i(t)$, $i=1,\ldots,n$, as functional observations rather than as high-dimensional discrete vectors, by assuming a basis representation of the curves as it has been already motivated, yielding
\begin{equation}\label{eq:xiB-spline}
    x_i(t) = \displaystyle \sum_{j=1}^{k+d} a_{ij} B_j(t) = \mathbf{B}(t)\mathbf{a}_i,
\end{equation}
where $\mathbf{a}_i\in \mathbb{R}^{k+d}$ is the vector of regression coefficients of the $i$-th functional observation $x_i(\cdot)$, and $\mathbf{B}(\cdot)=(B_1(\cdot),\ldots,B_{k+d}(\cdot))$ is the vector of B-spline basis functions. For simplicity of exposition, we assume the same basis $\mathbf{B}(\cdot)$ for all observations, although a different basis could be used for each $x_i(\cdot)$ if required. Following the approach in Section \ref{subsec:psplines}, each $\mathbf{a}_i$ is obtained as the solution to \eqref{eq:sosrPspline-matrix}, using as input data $\{(t_{m}, x_i(t_{m}))\}_{1\leq m\leq M}$. In other words, this is equivalent to fitting a smooth regression model with P-splines to each functional observation individually, which is why no common set of observation instants is needed.

Similarly, the functional regression coefficient $\beta(t)$ can be represented as
\begin{equation}\label{eq:sofbetaB-spline}
    \beta(t) = \displaystyle\sum_{j=1}^{k+d} b_j B_j(t) = \mathbf{B}(t)\mathbf{b},
\end{equation}
where $\mathbf{b}\in\mathbb{R}^{k+d}$ is the vector of regression coefficients to be estimated and $\mathbf{B}(\cdot)$ is defined as in \eqref{eq:xiB-spline}. Observe that, for the sake of simplicity,  the same B-spline basis is assumed in \eqref{eq:xiB-spline} and \eqref{eq:sofbetaB-spline}.
Then, the SOF linear regression model in \eqref{eq:sof} becomes
\begin{equation}\label{eq:sof-basis}
    \begin{array}{ll}
        y_i &= \displaystyle\int_0^T \mathbf{a}_i^\top \mathbf{B}(t)^\top \mathbf{B}(t) \mathbf{b}\,dt + \varepsilon_i = \\
        & = \mathbf{a}_i^\top \left(\displaystyle\int_0^T \mathbf{B}(t)^\top \mathbf{B}(t) \, dt \right) \mathbf{b} + \varepsilon_i = \\
        & = \mathbf{a}_i^\top \boldsymbol{\Psi} \mathbf{b} + \varepsilon_i,
    \end{array}
\end{equation}
where $\boldsymbol{\Psi} = \displaystyle\int_0^T \mathbf{B}(t)^\top \mathbf{B}(t)  dt\in\mathbb{R}^{(k+d)\times (k+d)}$ is the inner product matrix. Thus, we obtain a model of the same form as \eqref{eq:sosr} and \eqref{eq:sosrBspline}, and the vector of regression coefficients $\mathbf{b}$ can be estimated following the P-spline paradigm. In other words, the SOF linear regression model can be estimated using a P-spline approach by solving the penalized least-squares problem \eqref{eq:frm} where $\mathbf{u} = (y_1,\ldots,y_n)^\top,$ $\mathbf{Z} = \mathbf{A} \boldsymbol{\Psi},$  $\mathbf{A}\in\mathbb{R}^{n\times{(k+d)}}$ is the matrix whose $i$-th row is  $\mathbf{a}_i^\top$, $i=1,\ldots,n,$ and $\boldsymbol{\delta} = \mathbf{b}.$  The penalty term is defined as $P(\boldsymbol{\delta}) = \lambda \|\mathbf{D}\boldsymbol{\delta}\|_2^2$ as in \eqref{eq:sosrPspline-matrix}. 

\subsection{Function-on-scalar regression with P-splines \label{subsec:FOS}}

Let $\{(x_i, y_i(t))\}_{1\leq i\leq n}$ be a set of observations from a scalar covariate $X$ and a functional response variable $Y(t),$ with $t\in[0,T]\subset\mathbb{R}^+.$ Then, the function-on-scalar (FOS) linear regression model \citep[see, e.g.,][]{ramsay1997functional, ReissHuangMennes2010} is formulated as 
\begin{equation}\label{eq:fos}
     y_i(t) = \beta_0(t) + x_i\beta(t) + \varepsilon_i(t),
\end{equation}
where $\varepsilon_1,\ldots,\varepsilon_n$ are independent realizations of a zero-mean second-order stochastic process on $[0,T]$, with covariance function
\begin{equation}\label{eq:covfunction}
\sigma(t,s) \;=\; \mathrm{Cov}\left(\varepsilon_i(t),\varepsilon_i(s)\right),
\qquad t,s\in[0,T],\quad i=1,\ldots,n.
\end{equation}
In other words, error curves associated with different individuals are independent, but the errors within a curve need not be uncorrelated along the domain. We refer to the uncorrelated-errors case when $\sigma(t,s)=\sigma^2\,\mathbb{I}(t=s)$ for some $\sigma^2>0$, where $\mathbb{I}(\cdot)$ is the indicator function, equal to one if its argument holds and zero otherwise, and to the correlated-errors case otherwise. The latter is the rule rather than the exception for functional responses, which typically exhibit temporal correlation and heteroscedasticity across $t\in[0,T]$ \citep[see, e.g.,][]{goldsmith2011functional, ReissHuangMennes2010}. Assuming that the sample data are mean-centered at zero, we set $\beta_0(t)=0$, $t\in[0,T]$. The functional regression coefficient to be estimated is $\beta(t)$, $t\in[0,T]$.

As in the SOF linear regression model, we represent both $\beta(t)$ and $y_i(t)$ using B-spline basis expansions. Specifically, $\beta(t)$ is represented as in \eqref{eq:sofbetaB-spline}, and each functional response $y_i(t)$ is expressed as

\begin{equation}\label{eq:yiB-spline}
    y_i(t) = \displaystyle \sum_{j=1}^{k+d} c_{ij} B_j(t) = \mathbf{B}(t)\mathbf{c}_i,
\end{equation}
where $\mathbf{c}_i\in \mathbb{R}^{k+d}$ is the vector of regression coefficients for the $i-$th functional observation $y_i(\cdot)$ and $\mathbf{B}(\cdot)^\top\in\mathbb{R}^{k+d}$ is the vector of B-spline basis functions. Observe that, for the sake of simplicity, the same B-spline basis  is assumed in \eqref{eq:sofbetaB-spline} and \eqref{eq:yiB-spline}. As in the SOF case, each $\mathbf{c}_i$ is obtained as the solution to \eqref{eq:sosrPspline-matrix}, using as input data $\{(t_m, y_i(t_m))\}_{1\leq m\leq M}$, $i=1,\ldots,n$, where $t_m$ are the $M$ points at which the functional response variable has been observed. Again, we assume common observation instants for all individuals only for simplicity: as discussed in Section~\ref{subsec:SOF}, this is not required, and the instants may differ across individuals. Then, the FOS linear regression model in \eqref{eq:fos} becomes
\begin{equation}\label{eq:fos-basis}
    \mathbf{B}(t)\mathbf{c}_i = \mathbf{B}(t)\mathbf{b}x_i + \varepsilon_i(t),
\end{equation}
for $i=1,\ldots,n$ and $\varepsilon_i(t),$ $t\in[0,T],$ as in \eqref{eq:fos}. 

Because the response is functional, the least-squares criterion used to estimate $\mathbf{b}$ in \eqref{eq:fos-basis} corresponds to minimizing an integrated mean squared error (IMSE) over the domain $[0,T]$, rather than the standard mean squared error used for scalar responses in \eqref{eq:sosrPspline} or in the SOF linear regression model. 
Under uncorrelated errors, this leads to an equivalent matrix formulation of the optimization problem  in \eqref{eq:frm}, being
\begin{equation}\label{eq:uZfos}
\mathbf{u} = \vect\left(\mathbf{C}\boldsymbol{\Psi}^{1/2}\right)\in\mathbb{R}^{n(k+d)},
\,\,
\mathbf{Z} = \boldsymbol{\Psi}^{1/2} \otimes \mathbf{x}\in\mathbb{R}^{n(k+d)\times(k+d)},
\,\,
\boldsymbol{\delta} = \mathbf{b},
\end{equation}

\noindent where $\mathbf{C} \in \mathbb{R}^{n\times(k+d)}$ is the matrix whose $i$-th row is $\mathbf{c}_i^\top$,
$\mathbf{x} = (x_1,\ldots,x_n)^\top$ is the vector of scalar covariates, $\boldsymbol{\Psi}$ is the inner product matrix defined in \eqref{eq:sof-basis}, $\vect(\cdot)$ is the vectorization operator, and $\otimes$ denotes the Kronecker product. The penalty term is again given by $P(\boldsymbol{\delta}) = \lambda \|\mathbf{D}\boldsymbol{\delta}\|_2^2$,
as in \eqref{eq:sosrPspline-matrix}. The detailed derivation of these expressions can be found in Section \ref{subsec:app-fos-uncorrelated} of Appendix~\ref{app:FOS-errorsdep}.

Under correlated errors, the generalized least squares (GLS) framework has to be used instead of the ordinary least squares in order to account for the within-curve correlation.  The terms in \eqref{eq:uZfos} are the ones associated with an error process whose covariance operator is a multiple of the identity, and departures from that situation are accounted for by a single matrix $\boldsymbol{\Sigma}\in\mathbb{R}^{(k+d)\times(k+d)}$, which represents the covariance operator associated with $\sigma(t,s)$ in \eqref{eq:covfunction}. Since $\boldsymbol{\Sigma}$ is unknown, the model \eqref{eq:fos-basis} is estimated in two stages. In a first stage, the regression coefficients are estimated by solving \eqref{eq:frm} with \eqref{eq:uZfos}, i.e., under uncorrelated errors. In a second stage, an estimate $\widehat{\boldsymbol{\Sigma}}$ is computed from the resulting residual curves by functional principal component analysis (FPCA) \citep{yao2005functional}, and problem \eqref{eq:frm} is solved again with
\begin{equation}\label{eq:uZfos-gls}
\mathbf{u} = \vect\left(\mathbf{C}\boldsymbol{\Psi}^{1/2}\widehat{\boldsymbol{\Sigma}}^{-1/2}\right),
\qquad
\mathbf{Z} = \widehat{\boldsymbol{\Sigma}}^{-1/2} \boldsymbol{\Psi}^{1/2} \otimes \mathbf{x},
\qquad
\boldsymbol{\delta} = \mathbf{b},
\end{equation}
and the same penalty term. Note that the whole effect of the correlation structure is a re-weighting of $\mathbf{u}$ and $\mathbf{Z}$: the vector of regression coefficients and the penalty are left untouched, so both stages are instances of \eqref{eq:frm}. The detailed derivation of these expressions can be found in Section \ref{subsec:app-fos-correlated} of Appendix~\ref{app:FOS-errorsdep}.

\subsection{Function-on-function regression with P-splines \label{subsec:FOF}}

Let  $\{(x_i(s), y_i(t))\}_{1\leq i\leq n}$ be a set of observations from a functional covariate $X(s)$ and a functional response variable $Y(t),$ $s, t$ $\in[0,T]\subset \mathbb{R}^+$ without loss of generality. Then, the function-on-function (FOF) linear regression model \citep[see, e.g.,][]{ivanescu2015penalized,ramsay1997functional} is stated as
\begin{equation}\label{eq:fof}
     y_i(t) = \beta_0(t) + \int_0^T x_i(s)\beta(s,t)\,ds + \varepsilon_i(t),
\end{equation}
where the error functions $\varepsilon_i:[0,T]\longrightarrow \mathbb{R}$ are assumed to behave as those in~\eqref{eq:fos}, i.e., they are independent across individuals with covariance function $\sigma(t,s)$ as in~\eqref{eq:covfunction}. Assuming that the sample data is mean-centered at zero, $\beta_0(t) = 0,$ $ t\in[0,T]$, and the bivariate regression coefficient to be estimated is the surface $\beta(s,t),$ $s,t~\in~[0,T].$

We first address the estimation of the general model \eqref{eq:fof} in Section~\ref{subsubsec:generalFOF}. Section~\ref{subsubsec:concurrent} then considers the so-called concurrent model, in which the response at time $t$ depends on the covariate only through its value at that same time point. This model is widely used in practice, as it is simpler and easier to interpret than the general model, and it fits many applications in which the covariate acts on the response at the same time instant. Although it can be seen as a limiting case of \eqref{eq:fof}, its estimation does not follow from that of the general model and requires a separate treatment, as discussed there. In both cases, estimation is first described under uncorrelated errors; the two-stage GLS strategy that accounts for correlated errors, analogous to the one used for the FOS linear regression model in Section~\ref{subsec:FOS}, is presented in Section~\ref{subsubsec:FOF-dependent}.

\subsubsection{The general model \label{subsubsec:generalFOF}}

As in the SOF and FOS linear regression models, the coefficient surface $\beta(s,t)$ is represented using B-spline basis functions,  now through a tensor-product expansion. Without loss of generality, we assume that in both dimensions the number of intervals in which the domain $[0,T]$ is split is~$k,$ and the degree of the basis functions is~$d$, yielding:
\begin{equation}\label{eq:fofbetaB-spline}
    \beta(s,t) = \displaystyle\sum_{j=1}^{k+d} \sum_{\ell=1}^{k+d} b_{j\ell} B_j(t) \Gamma_\ell(s)= \mathbf{B}(t)\mathcal{B}\boldsymbol{\Gamma}(s)^\top,
\end{equation}
where $\mathbf{B}(\cdot)^\top, \boldsymbol{\Gamma}(\cdot)^\top\in\mathbb{R}^{k+d}$ are the vectors of basis functions in the $t-$ and $s-$directions, respectively, and $\mathcal{B}=(b_{j\ell})\in\mathbb{R}^{(k+d)\times(k+d)}$ is the matrix of regression coefficients to be estimated. 

Assuming also the basis representations of the covariate and the response,
$x_i(s)=\boldsymbol{\Gamma}(s)\mathbf{a}_i$ and $y_i(t)=\mathbf{B}(t)\mathbf{c}_i$ as in
\eqref{eq:xiB-spline} and \eqref{eq:yiB-spline}, the FOF linear regression model in \eqref{eq:fof} becomes:
\begin{equation}\label{eq:fof-basis}
    \begin{array}{ll}
        \mathbf{B}(t)\mathbf{c}_i &= \displaystyle\int_0^T x_i(s) \mathbf{B}(t) \mathcal{B} \boldsymbol{\Gamma}(s)^\top ds + \varepsilon_i(t) = \\
        & =  \mathbf{B}(t) \mathcal{B}\mathbf{w}_i + \varepsilon_i(t),
    \end{array}
\end{equation}
where $\mathbf{w}_i = \displaystyle\int_0^T \boldsymbol{\Gamma}(s)^\top x_i(s) ds \in \mathbb{R}^{k+d}$, $i=1,\ldots,n.$ Since  the functional covariate has also a basis representation,  
\begin{equation}\label{eq:wi}
    \mathbf{w}_i = \int_0^T \boldsymbol{\Gamma}(s)^\top \boldsymbol{\Gamma}(s)\, ds \;\mathbf{a}_i = \boldsymbol{\Psi}_{\Gamma}\,\mathbf{a}_i,
\end{equation}
that is, $\mathbf{w}_i$ is obtained in closed form from the coefficient vector $\mathbf{a}_i$ of the $i$-th functional covariate and the inner product matrix $\boldsymbol{\Psi}_{\Gamma} = \int_0^T \boldsymbol{\Gamma}(s)^\top \boldsymbol{\Gamma}(s)\, ds \in \mathbb{R}^{(k+d)\times(k+d)}$, exactly as in the SOF linear regression model \eqref{eq:sof-basis}. In particular, if the same basis is used in both dimensions, namely $\boldsymbol{\Gamma} = \mathbf{B}$, then $\boldsymbol{\Psi}_{\Gamma}\equiv \boldsymbol{\Psi}$.

Note that \eqref{eq:fof-basis} has exactly the structure of the FOS linear regression model \eqref{eq:fos-basis}, with the scalar covariate $x_i$ replaced by the vector $\mathbf{w}_i$ and the coefficient vector $\mathbf{b}$ replaced by the coefficient matrix $\mathcal{B}$. Consequently, the estimation of $\mathcal{B}$ proceeds as that of $\mathbf{b}$ in \eqref{eq:fos-basis}: the least-squares criterion corresponds to minimizing an IMSE over $[0,T]$,  which leads to an optimization problem of the form given in~\eqref{eq:frm} with
\begin{equation}\label{eq:uZfof}
\begin{gathered}
\mathbf{u} = \vect\left(\mathbf{C}\boldsymbol{\Psi}^{1/2}\right)\in\mathbb{R}^{n(k+d)}, \qquad
\mathbf{Z} = \boldsymbol{\Psi}^{1/2} \otimes \mathbf{W}\in\mathbb{R}^{n(k+d)\times (k+d)^2},\\
\boldsymbol{\delta} = \vect(\mathcal{B})\in\mathbb{R}^{(k+d)^2},
\end{gathered}
\end{equation}
where $\mathbf{C}, \mathbf{W} \in \mathbb{R}^{n\times(k+d)}$ contain, respectively, the coefficient
vectors $\mathbf{c}_i$ and $\mathbf{w}_i$ in their $i-$th row, $i=1,\ldots,n$, and
$\boldsymbol{\Psi} = \int_0^T \mathbf{B}(t)^\top \mathbf{B}(t)\,dt \in
\mathbb{R}^{(k+d)\times (k+d)}$. In view of \eqref{eq:wi},
$\mathbf{W} = \mathbf{A}\boldsymbol{\Psi}_{\Gamma}$, where $\mathbf{A}\in\mathbb{R}^{n\times(k+d)}$ is
the matrix whose $i$-th row is $\mathbf{a}_i^\top$, as in the SOF linear regression model. The details of these
calculations are provided in Appendix~\ref{app:FOF-general}. 

Since the functional regression coefficient is now a surface,  the penalty term $P(\boldsymbol{\delta})$ in \eqref{eq:frm}  is the anisotropic tensor-product difference penalty of bivariate P-splines \citep{currie2006generalized,eilers2003multivariate}, which controls smoothness separately in each direction: 
\begin{equation}\label{eq:penalty2d}
     P(\boldsymbol{\delta}) =
    \lambda_t \|(\mathbf{D} \otimes I_{k+d})\boldsymbol{\delta}\|_2^2
    + \lambda_s \|(I_{k+d} \otimes \mathbf{D})\boldsymbol{\delta}\|_2^2,
\end{equation}
where $\mathbf{D}$ is the second-order difference matrix of size $(k+d-2)\times(k+d)$,
$I_{k+d}$ is the identity matrix of order $k+d$, and $\lambda_s, \lambda_t \ge 0$ are smoothing parameters controlling the roughness of $\beta(s,t)$ along the $s$- and $t$-directions, so $\mathbf{D}\otimes I_{k+d}$ differences along $t$ and $I_{k+d}\otimes\mathbf{D}$ differences along $s$.

\subsubsection{The concurrent model \label{subsubsec:concurrent}}

A special case of the FOF linear regression model \eqref{eq:fof} is the so-called concurrent model, in which the response at time $t$ depends only on the value of the predictor at the same time point. Formally, this corresponds to a regression coefficient surface that concentrates all its mass on the diagonal $s=t$,
\begin{equation}\label{eq:betaconcurrent}
    \beta(s,t) = \beta(t)\,\mathbb{I}(s=t),
\end{equation}
where $\mathbb{I}(s=t)$ denotes the indicator function that is equal to $1$ when $s=t$ and $0$ otherwise. Substituting \eqref{eq:betaconcurrent} into \eqref{eq:fof}, the integral reduces to a point-wise multiplication, and the FOF linear regression model becomes
\begin{equation}\label{eq:fofconcurrent}
    y_i(t) = x_i(t)\,\beta(t) + \varepsilon_i(t), \qquad i=1,\ldots,n,
\end{equation}
where the functional regression coefficient to be estimated is now the univariate function $\beta(t)$, $t\in[0,T]$, represented as in \eqref{eq:sofbetaB-spline}. 

To estimate $\beta(t)$ in \eqref{eq:fofconcurrent}, the formulations of the concurrent model that we are aware of \citep[see, e.g.,][]{ghosal2023shape, ivanescu2015penalized,PenFFR} are written in terms of the response evaluated at the observation points $t_m\in[0,T]$, $m=1,\ldots,M$: the coefficient function (and, in some cases, the covariates) is represented in a basis, but the response is handled through its raw discrete observations. Here, instead, and consistently with the SOF and FOS linear regression models above, we keep the basis representations of the covariate and the response, $x_i(t)=\mathbf{B}(t)\mathbf{a}_i$ and $y_i(t)=\mathbf{B}(t)\mathbf{c}_i$ as in \eqref{eq:xiB-spline} and \eqref{eq:yiB-spline}, and estimate $\mathbf{b}$ by minimizing the corresponding IMSE, whose integrals are computed exactly. As shown in Remark~\ref{rem:rawdata} below, the raw-data formulation is recovered as a quadrature approximation of this criterion.

Unlike in the SOF and FOS settings, the fitted term $x_i(t)\beta(t)$ in \eqref{eq:fofconcurrent} is a product of two splines, and such a product is no longer a spline on the same basis: it is a piecewise polynomial of degree $2d$. The coefficients of the response and those of the fitted term can therefore not be compared term by term in the original basis. As shown in Appendix~\ref{app:FOF-concurrent}, this is overcome by writing both terms in the tensor-product family $\{B_j(t)B_\ell(t)\}_{1\leq j,\ell\leq k+d}$, whose row vector is $\mathbf{B}(t)\otimes\mathbf{B}(t)$, lifting the response to this family through the partition-of-unity property of B-splines. This yields once again an optimization problem of the form
\eqref{eq:frm}, with

\begin{equation}\label{eq:uZconcurrent}
\begin{gathered}
\mathbf{u}
= \begin{pmatrix}
\bar{\boldsymbol{\Psi}}^{1/2}\left(\mathbf{1}\otimes\mathbf{c}_1\right)\\
\vdots\\
\bar{\boldsymbol{\Psi}}^{1/2}\left(\mathbf{1}\otimes\mathbf{c}_n\right)
\end{pmatrix}\in\mathbb{R}^{n(k+d)^2},
\quad
\mathbf{Z}
= \begin{pmatrix}
\bar{\boldsymbol{\Psi}}^{1/2}\left(\mathbf{a}_1\otimes I_{k+d}\right)\\
\vdots\\
\bar{\boldsymbol{\Psi}}^{1/2}\left(\mathbf{a}_n\otimes I_{k+d}\right)
\end{pmatrix}\in\mathbb{R}^{n(k+d)^2\times(k+d)},\\[1ex]
\boldsymbol{\delta}=\mathbf{b}\in\mathbb{R}^{k+d},
\end{gathered}
\end{equation}
where $\mathbf{1}\in\mathbb{R}^{k+d}$ is the vector of ones, and $\bar{\boldsymbol{\Psi}}$ plays for that larger family the role that $\boldsymbol{\Psi}$ plays for $\mathbf{B}(t)$ (its expression and its numerical evaluation are given in Appendix~\ref{app:FOF-concurrent}). The penalty term is again $P(\boldsymbol{\delta}) = \lambda \|\mathbf{D}\boldsymbol{\delta}\|_2^2$, as in \eqref{eq:sosrPspline-matrix}. Note that this change of representation affects only $\mathbf{u}$ and $\mathbf{Z}$, and never the vector of unknowns: the coefficient function retains its B-spline representation $\beta(t)=\mathbf{B}(t)\mathbf{b}$ throughout.

\begin{remark}[Connection with the raw-data formulation]\label{rem:rawdata}
Straightforward algebra shows that the two matrices that determine the least-squares term of \eqref{eq:frm}, with $\mathbf{u}$ and $\mathbf{Z}$ as in \eqref{eq:uZconcurrent}, admit the closed forms 
\begin{equation}\label{eq:normalconcurrent}
\mathbf{Z}^{\top} \mathbf{Z}
= \sum_{i=1}^{n}\int_0^T x_i(t)^2\,\mathbf{B}(t)^{\top}\mathbf{B}(t)\,dt,
\,\,
\mathbf{Z}^{\top}\mathbf{u}
= \sum_{i=1}^{n}\int_0^T x_i(t)\,y_i(t)\,\mathbf{B}(t)^{\top}\,dt.
\end{equation}
Replacing the integrals in \eqref{eq:normalconcurrent} by sums over the observation grid $\{t_m\}_{1\le m\le M}$ recovers the classical raw-data formulation of the concurrent model, $\mathbf{y}_i=\mathrm{diag}(\mathbf{x}_i)\,\mathbf{B}\,\mathbf{b}+\boldsymbol{\varepsilon}_i$, where $\mathbf{y}_i=(y_i(t_1),\ldots,y_i(t_M))^{\top}$, $\mathrm{diag}(\mathbf{x}_i)$ is the diagonal matrix with $\mathbf{x}_i=(x_i(t_1),\ldots,x_i(t_M))^{\top}$ on its diagonal, and $\mathbf{B}=(\mathbf{B}(t_1)^{\top},\ldots,\mathbf{B}(t_M)^{\top})^{\top}$, which corresponds to solving \eqref{eq:frm} with $\mathbf{u} = (\mathbf{y}_1^\top,\ldots,\mathbf{y}_n^\top)^\top$ and $\mathbf{Z} = (\mathrm{diag}(\mathbf{x}_1)\mathbf{B},\ldots,\mathrm{diag}(\mathbf{x}_n)\mathbf{B})^\top$. In this sense, the raw-data approach can be viewed as a quadrature approximation of the proposed  estimation procedure in this paper,  in which the integrals are computed exactly.
\end{remark}

\subsubsection{Estimation under correlated errors \label{subsubsec:FOF-dependent}}

The estimation of the FOF linear regression model, both in the general and the concurrent case, has been described so far under the assumption of uncorrelated errors. When $\sigma(t,s)\neq \sigma^2\mathbb{I}(t=s)$, the same two-stage GLS strategy described for the FOS linear regression model in Section~\ref{subsec:FOS} is used, the only difference being that the first stage now estimates an FOF linear regression model under uncorrelated errors. Once the covariance matrix $\boldsymbol{\Sigma}$ has been estimated from the residual curves by FPCA
\citep{yao2005functional}, the second stage solves \eqref{eq:frm} with
\begin{equation}\label{eq:uZfof-gls}
\mathbf{u} = \vect\left(\mathbf{C}\boldsymbol{\Psi}^{1/2}\widehat{\boldsymbol{\Sigma}}^{-1/2}\right),
\qquad
\mathbf{Z} = \widehat{\boldsymbol{\Sigma}}^{-1/2}\boldsymbol{\Psi}^{1/2}\otimes\mathbf{W},
\qquad
\boldsymbol{\delta}=\vect(\mathcal{B}^\top),
\end{equation}
for the general model, and with the analogous re-weighting of \eqref{eq:uZconcurrent} for the concurrent one. The details are given in Appendices~\ref{app:FOF-general} and
\ref{app:FOF-concurrent}.

Note that \eqref{eq:uZfof-gls} is exactly the expression obtained for the FOS linear regression model in
Section~\ref{subsec:FOS}, with the vector of scalar covariates $\mathbf{x}$ replaced by the matrix $\mathbf{W},$ and $\mathbf{b}$ by $\vect(\mathcal{B}^\top)$. The penalty term is left unchanged, i.e., \eqref{eq:penalty2d} for the general model and $\lambda\|\mathbf{D}\boldsymbol{\delta}\|_2^2$ for the concurrent one, so both stages are instances of \eqref{eq:frm}.

\section{FunSCOPE: Shape-constrained functional regression with P-splines \label{sec:Shape-FR}}

The P-spline estimators of Section~\ref{sec:FRPsplines} do not guarantee that the estimated functional regression coefficient satisfies the shape requirements dictated by the nature of the problem or by expert knowledge. We address this limitation by developing a functional shape-constrained conic P-spline estimator 
(FunSCOPE) which combines two elements. First, the functional covariate and/or response are represented through the B-spline bases as described in Section~\ref{sec:FRPsplines}, so that the data are treated as functional objects. Second, we extend the conic optimization approaches of \cite{Navarro2023AMC} and \cite{navarro2024} to the functional framework, which allows any combination of the shapes considered to be imposed on the functional coefficient through conditions on its spline coefficients. For univariate functional coefficients, namely those in the SOF, FOS, and concurrent FOF linear regression models, these conditions are both necessary and sufficient for the shapes to hold over the whole domain. Section~\ref{subsec:cons-uni} describes this univariate case, corresponding to the shape-constrained estimation of $\beta(t)$ in \eqref{eq:sof}, \eqref{eq:fos}, and \eqref{eq:fofconcurrent}. Section~\ref{subsec:cons-bi} presents the bivariate case, corresponding to $\beta(t,s)$ in \eqref{eq:fof}, where the conditions are neither necessary nor sufficient in the whole domain, but they are on some curves belonging to it.

\subsection{Shape-constrained estimation of \texorpdfstring{$\beta(t)$}{beta(t)}\label{subsec:cons-uni}}

The  SOF, FOS, and concurrent FOF linear regression models in \eqref{eq:sof}, \eqref{eq:fos}, and \eqref{eq:fofconcurrent}, respectively, lead to different expressions for the response vector $\mathbf{u}$ and the design matrix $\mathbf{Z}$ in~\eqref{eq:frm}, while the regression coefficients $\boldsymbol{\delta}$ remain equal to the coefficients of the basis representation of $\beta(t)$, namely $\boldsymbol{\delta} = \mathbf{b}$ in all three cases. In particular, $\mathbf{Z}$ is not, in general, a matrix of evaluations of B-spline basis functions, as in standard smooth regression (see Section \ref{subsec:psplines}), thus raising the question of whether the shape-constrained methodology of \cite{Navarro2023AMC} still applies in this setting.

\citet{Navarro2023AMC} address the problem of estimating smooth regression models with P-splines as described in Section \ref{subsec:psplines} with shape constraints. To incorporate requirements on the sign, monotonicity, and/or curvature into the fitting procedure, a conic optimization approach is developed that provides necessary and sufficient conditions for estimating a curve subject to such constraints, for an arbitrary basis degree $d$. The key point is that these conditions are imposed as constraints on the  unknown regression coefficients, i.e., $\boldsymbol{\theta}$ in \eqref{eq:sosrPspline-matrix} (or $\boldsymbol{\delta}= \boldsymbol{\theta}$ in \eqref{eq:frm}) rather than on the fitted values.

For all the functional regression models considered here, the coefficient function $\beta(t)$ retains its B-spline representation as in \eqref{eq:sofbetaB-spline}. The transformation of the functional regression model into~\eqref{eq:frm} changes the data-fitting term through $\mathbf{u}$ and $\mathbf{Z}$, but it does not alter this representation of $\beta(t)$ or the corresponding penalty. In other words,  $\boldsymbol{\delta}= \mathbf{b}$ in all the three cases. Consequently, the characterization of sign, monotonicity, and curvature constraints developed by \citet{Navarro2023AMC} remains applicable to $\beta(t)$. These constraints depend only on the B-spline representation of the coefficient function $\beta(t)$, and therefore on its coefficient vector $\boldsymbol{\delta}$($= \mathbf{b}$), and not on whether the design matrix $\mathbf{Z}$ itself is a B-spline evaluation matrix. Therefore, the SOF, FOS, and concurrent FOF linear regression models in \eqref{eq:sof}, \eqref{eq:fos}, and \eqref{eq:fofconcurrent}, respectively, can be
estimated under shape constraints by combining their model-specific
least-squares term ($\|\mathbf{u}-\mathbf{Z}\boldsymbol{\delta}\|_2^2$) and penalty ($P(\boldsymbol{\delta})$) with the corresponding constraints on $\beta(t)$, as defined in \eqref{eq:sofbetaB-spline}.

In the following, we explain how to impose that $\beta(t)$ is non-negative for $t\in[0,T].$ Let us consider a B-spline basis representation of  $\beta(t)$   as in \eqref{eq:sofbetaB-spline} for $t\in[0,T],$ which yields a univariate piecewise polynomial of degree $d$. Restricting it to the interval  $[\tau_q, \tau_{q+1}],$ it can be expressed as:
\begin{equation}\label{eq:beta-polynomial}
\left. \beta(t)\right|_{t\in[\tau_q,\tau_{q+1}]} = \sum_{r=0}^d\left(\sum_{p=q-d}^q\delta_p h_{pr}^{(q)}\right)t^r,
\end{equation}
where $\displaystyle \sum_{p=q-d}^q \delta_p h_{pr}^{(q)}$, for $r = 0, \dots, d$, is the degree-$r$ coefficient, obtained as a linear combination (via $\delta_p$) of the $d+1$ non-zero polynomial pieces on $[\tau_q, \tau_{q+1}]$ (via $h_{pr}^{(q)}$), one from each B-spline basis function active on that interval ($q=d+1,\ldots,k+d)$.

\cite{bertsimas2002relation} provide a characterization of a univariate polynomial being non-negative on a closed real interval in terms of the existence of a positive semidefinite matrix of dimension $d+1$ satisfying a set of linear constraints involving its elements, the coefficients of the polynomial, and the bounds of the interval on which it is defined. In particular, these conditions are necessary and sufficient for the polynomial to be non-negative on the interval. Applying this characterization to $\left. \beta(t)\right|_{t\in[\tau_q,\tau_{q+1}]}$ as in \eqref{eq:beta-polynomial} for $q=d+1,\ldots,k+d,$ the vector of regression coefficients $\boldsymbol{\delta}$ is estimated by solving the following conic optimization problem:
\begingroup\fontsize{9pt}{13.6pt}\selectfont
\allowdisplaybreaks
\begin{subequations}
\label{eq:constr-beta-t}
\begin{alignat}{2}
\min \quad & \|\mathbf{u} - \mathbf{Z}\boldsymbol{\delta}\|_2^2 + \lambda\|\mathbf{D}\boldsymbol{\delta}\|_2^2  \label{eq:constr-beta-t-obj} \\
\text{s.t.} \quad & \sum_{\substack{i,j\in\{1,\ldots,d+1\} \\ i+j = 2\ell +1}} \xi_{ij}^{(q)} = 0, \quad  \substack{\ell = 1, \ldots, d,\\ q=d+1,\ldots,k+d} \label{eq:constr-beta-t-cons0} \\
 & \sum_{\substack{i,j\in\{1,\ldots,d+1\} \\ i+j = 2(\ell +1)}} \xi_{ij}^{(q)}= \sum_{m=0}^\ell \sum_{r=m}^{d+m-\ell} \left(\sum_{p = q-d}^q \delta_p h_{pr}^{(q)}  \right) \binom{r}{m} \binom{d-r}{\ell - m} \tau_q^{r-m}\tau_{q+1}^m , \quad  \substack{\ell = 0, \ldots, d,\\ q=d+1,\ldots,k+d} \label{eq:constr-beta-t-consCOEF} \\
 & \mathbf{\Xi}^{(q)} = \left(\xi_{ij}^{(q)}\right)_{i,j=1,\ldots,d+1} \in \mathcal{S}^{d+1}_+, \quad \substack{q=d+1,\ldots,k+d} \label{eq:constr-beta-t-consSDP} \\
 & \boldsymbol{\delta} = (\delta_1,\ldots,\delta_{k+d})^\top \in \mathbb{R}^{k+d} \label{eq:constr-beta-t-consdelta}
\end{alignat}
\end{subequations}
\endgroup

\noindent where $\binom{N}{M}$ refers to the binomial coefficient of $N$ and $M$, and $\mathcal{S}^{d+1}_+$ stands for the cone of positive semidefinite matrices of dimension $d+1.$ On the one hand, we point out that \eqref{eq:constr-beta-t-obj} refers to the penalized least-squares criterion in SOF, FOS and concurrent FOF linear regression models in \eqref{eq:sof}, \eqref{eq:fos}, and \eqref{eq:fofconcurrent}, respectively. On the other hand, constraints \eqref{eq:constr-beta-t-cons0}--\eqref{eq:constr-beta-t-consSDP} ensure that the functional regression coefficient $\beta(t)$ in all those three cases is non-negative in the whole domain $[0,T]$. Finally, \eqref{eq:constr-beta-t-consdelta} defines the dimension of the regression coefficients $\boldsymbol{\delta}$, which are the decision variables of the problem jointly with the matrices $\mathbf{\Xi}^{(q)},$ $q=d+1,\ldots,k+d.$

Observe that sign, monotonicity, or curvature requirements on $\beta(t)$ follow straightforwardly from \eqref{eq:constr-beta-t} by imposing the corresponding non-negativity constraints on $\beta(t)$, $\beta'(t)$, or $\beta''(t)$ (or on their negatives, for non-positivity, decrease, or concavity, respectively). Furthermore, these different conditions can be combined by incorporating the corresponding set of constraints into the resulting optimization problem. Finally, we highlight that different shape constraints can also be imposed on subintervals within $[0,T]$, by simply considering \eqref{eq:constr-beta-t-cons0}--\eqref{eq:constr-beta-t-consSDP}, or their counterparts for other shapes, over the corresponding subset of indices within $q = d+1, \ldots, k+d$.

\subsection{Shape-constrained estimation of \texorpdfstring{$\beta(s,t)$}{beta(s,t)}\label{subsec:cons-bi}}

The general FOF linear regression model detailed in  Section \ref{subsubsec:generalFOF} can also be expressed as a penalized least-squares problem of the form in \eqref{eq:frm}. In this case, however, the vector of regression coefficients $\boldsymbol{\delta}$ arises from vectorizing the coefficient matrix that defines the bivariate functional regression coefficient  $\beta(s,t)$ in \eqref{eq:fofbetaB-spline}, yielding a bivariate piecewise polynomial. Unlike the univariate case, where necessary and sufficient conditions for nonnegativity (and related shape constraints) hold regardless of the polynomial degree, no such general result is available for bivariate polynomials. Consequently, exact necessary and sufficient conditions for shape constraints on $\beta(s,t)$ cannot be guaranteed globally for the general FOF case.
Rather than leaving this issue unaddressed, we follow the approach of \citet{navarro2024}, which extends the conic optimization framework to the general FOF linear regression model by imposing the constraints on a finite set of curves of the surface.

\citet{navarro2024} address the problem of estimating shape-constrained multivariate smooth regression models by imposing shape constraints on a finite set of curves lying on the hypersurface defined by the function being estimated, following the univariate approach of \citet{Navarro2023AMC}. Specifically, they consider a skeleton of curves whose projections onto each dimension coincide with the position of the knots, and apply the method of \citet{Navarro2023AMC} to each curve individually. This allows different shape constraints (sign, monotonicity and curvature) to be imposed, and combined, along each dimension. Because these constraints are only enforced on the skeleton, this approach does not guarantee that the conditions hold in the regions between curves. However, as these authors report, the smoothness imposed by the underlying P-spline estimation prevents serious deviations, and the shape requirements usually hold in practice.

Similarly to the smooth regression case described above, the shape-constrained methodology of \citet{navarro2024} remains applicable to the estimation of 
$\beta(s,t)$, since it retains its B-spline representation, i.e., $\boldsymbol{\delta}=\vect(\mathcal{B}),$ as in \eqref{eq:fofbetaB-spline}. Then, as in the SOF, FOS, and concurrent FOF linear regression models, $\beta(s,t)$ can be estimated under shape constraints by combining its model-specific least-squares term (see \eqref{eq:uZfof}) and penalty term (see \eqref{eq:penalty2d}) with the constraints on the curves whose projections onto the $t-$ and $s-$axes coincide with the positions of the knots, respectively. 

Technically, holding $t$ fixed at a given value, the surface described by $\beta(s,t)$ reduces to a univariate piecewise polynomial of degree $d$ in $s$. In particular, if $t$ is set equal to each knot $\tau_q$, $q = d+1, \ldots, k+d$, while $s$ varies over $[0,T]$, then $\beta(s,\tau_q)$ defines a curve, and the collection of these curves forms a skeleton of the surface in the $s$-dimension. Conversely, since the same knot sequence is used in both dimensions, as assumed in Section \ref{subsubsec:generalFOF}, fixing $s$ equal to each knot $\tau_q$ while $t$ varies over $[0,T]$ yields the curves $\beta(\tau_q,t)$, which analogously form a skeleton of the surface in the $t$-dimension.  Since each of these curves is a univariate piecewise polynomial, the necessary and sufficient conditions of the univariate case apply to it exactly. Assuming that non-negativity of $\beta(s,t)$ must be imposed via a separate set of constraints for each of the two families of skeleton curves, the conic optimization problem to be solved is as follows:

\begin{small}
\begin{subequations}
\label{eq:constr-beta-st}
\begin{alignat}{2}
\min \quad & \|\mathbf{u} - \mathbf{Z}\boldsymbol{\delta}\|_2^2 + \lambda_t \|(\mathbf{D} \otimes I_{k+d})\boldsymbol{\delta}\|_2^2
    + \lambda_s \|(I_{k+d} \otimes \mathbf{D})\boldsymbol{\delta}\|_2^2  \label{eq:constr-beta-st-obj} \\
\text{s.t.} \quad & \beta(s,\tau_q) \text{ verifies \eqref{eq:constr-beta-t-cons0}--\eqref{eq:constr-beta-t-consSDP}, }\, s\in[0,T],\,  q=d+1,\ldots,k+d\label{eq:constr-beta-st-t} \\
 & \beta(\tau_q,t) \text{ verifies \eqref{eq:constr-beta-t-cons0}--\eqref{eq:constr-beta-t-consSDP}, }\, t\in[0,T],\,  q=d+1,\ldots,k+d \label{eq:constr-beta-st-s} \\
 & \boldsymbol{\delta}  \in \mathbb{R}^{(k+d)^2}. \label{eq:constr-beta-st-consdelta}
\end{alignat}
\end{subequations}
\end{small}

The objective function in \eqref{eq:constr-beta-st-obj} is the penalized least-squares criterion, with $\mathbf{u}$, $\mathbf{Z}$ and $\boldsymbol{\delta}$ as in \eqref{eq:uZfof} and the penalty term as in \eqref{eq:penalty2d}. Constraints \eqref{eq:constr-beta-st-t} (resp. \eqref{eq:constr-beta-st-s})  impose non-negativity on the curves defined over $s-$dimension (resp. $t-$dimension) by following the univariate approach stated in problem \eqref{eq:constr-beta-t}. Finally, \eqref{eq:constr-beta-st-consdelta} defines the dimension of the regression coefficients $\boldsymbol{\delta}.$

The same concluding remarks as in Section \ref{subsec:cons-uni} apply here. First, other shape constraints follow from \eqref{eq:constr-beta-st} by imposing the corresponding constraints on the curves, or their derivatives, in each dimension; furthermore, these different constraints can be combined within each dimension by incorporating the corresponding sets of constraints into the resulting optimization problem. Second, shape constraints can also be imposed on subintervals within 
$[0,T]$, in each dimension. Finally, we point out that we have assumed equal knot sequences in the $t$- and $s$-dimensions for simplicity of exposition, but our methodology still holds if they are different.

\section{Computational experiments}
\label{Computational experiments}

The aim of this section is to illustrate the performance of the proposed approach. Section \ref{Setup} describes the details of the experimental setup.  Sections \ref{Results SOFR}, \ref{Results FOSR} and \ref{Results FOFR} present the results for SOF, FOS and concurrent FOF linear regression models, respectively. 
Finally, Section~\ref{subsec:summary} summarizes the main conclusions derived from this analysis.

\subsection{Setup}
\label{Setup}

\paragraph{Compared methods.}
The performance of FunSCOPE for the different functional regression models is assessed on simulated datasets inspired by the setting proposed in \citet{ghosal2023shape}, which allows for a controlled comparison of the methodologies under known data-generating conditions. We compare three approaches: FunSCOPE, the solution provided by our approach when the shape constraints of each scenario are imposed; FunSCOPE-U, its unconstrained counterpart, obtained by solving \eqref{eq:frm} as it is, namely without any shape constraints; and the approach of \citet{ghosal2023shape}. FunSCOPE and the approach of \citet{ghosal2023shape} are evaluated under the same shape constraints in each scenario. The approach of \citet{han2025inference} is not included in the comparison, since it only addresses FOS regression and its open-source implementation does not readily accommodate the shape constraints of our FOS scenario.

\paragraph{Data generation.}
The simulation scenarios used to generate the data for each functional regression model are described in the corresponding sections below. Common to all scenarios, the functional data (the response, the covariate, or both) are observed on a grid of $m=50$ equally spaced instants $t_1,\ldots,t_m$ within the corresponding domain. For each scenario, three sample sizes are considered, $n\in\{25,50,100\}$, and each simulation setting is replicated 200 times.

\paragraph{Implementation details.}
In FunSCOPE and FunSCOPE-U, all functional objects are represented with cubic B-splines ($d=3$) on $k=14$ equally spaced intervals, i.e., $k+d=17$ basis functions, with a second-order difference penalty. These values are fixed in advance for all scenarios and sample sizes, as the basis is rich enough and the penalty controls smoothness. The smoothing parameter $\lambda$ is selected by minimizing the generalized cross-validation (GCV) criterion of the unconstrained problem, and the same value is then used in the shape-constrained one, so that FunSCOPE and FunSCOPE-U differ only in the constraints. In the FOS and FOF scenarios, where the errors are correlated, both are fitted with the two-stage GLS strategy described in Sections~\ref{subsec:FOS} and~\ref{subsubsec:FOF-dependent}.

\paragraph{Performance measures.}
We evaluate the methods in terms of prediction error, recovery of the functional regression coefficient, and computing time.

\begin{itemize}
    \item \emph{Prediction error.} In the SOF model, the response is scalar. Given the vector of observed responses $\mathbf{y}=(y_1,\ldots,y_n)^\top$ and the predictions $\hat{\mathbf{y}}=(\hat{y}_1,\ldots,\hat{y}_n)^\top$ of the estimated model, we use the Mean Squared Error (MSE):
\begin{equation}\label{eq:mse-response}
\mathrm{MSE}(\hat{\mathbf{y}}) = \frac{1}{n}\sum_{i=1}^{n} (y_i - \hat{y}_i)^2.
\end{equation}
In the FOS and FOF models, the response is a function, so for each observed curve $y_i(t)$ and predicted one $\hat{y}_i(t)$, $t\in\mathcal{T}$ and $i=1,\ldots,n$, where $\mathcal{T}$ denotes the domain interval of the functional data in each scenario and $|\mathcal{T}|$ its length, the squared error is integrated over the domain and then averaged over the $n$ observations, giving the Integrated Mean Squared Error (IMSE):
\begin{equation}\label{eq:imse-response}
\mathrm{IMSE}(\hat{y}) = \frac{1}{n}\sum_{i=1}^{n} \frac{1}{|\mathcal{T}|}\int_{\mathcal{T}} \left( y_i(t) - \hat{y}_i(t) \right)^2 dt.
\end{equation}
In FunSCOPE (resp. FunSCOPE-U), functional responses are represented through B-spline basis expansions, so $y_i(t)$ in \eqref{eq:imse-response} is the pre-smoothed curve, whereas for \citet{ghosal2023shape}, whose method works directly with the raw data, it is the observed curve at the instants $t_1,\ldots,t_m$. Consequently, the $\mathrm{IMSE}(\hat{y})$ values of FunSCOPE (resp. FunSCOPE-U) and  \citet{ghosal2023shape} are not fully comparable.

 \item \emph{Coefficient recovery.} For all three methods and all models, we measure how well the functional regression coefficient is recovered through the IMSE of the estimated functional regression coefficient $\hat{\beta}(t)$ with respect to the true one $\beta(t)$:
\begin{equation}\label{eq:imse-beta}
\mathrm{IMSE}(\hat{\beta}) = \frac{1}{|\mathcal{T}|}\int_{\mathcal{T}} \left( \beta(t) - \hat{\beta}(t) \right)^2 dt.
\end{equation}
This is the measure of main interest for our purposes, since good predictive performance does not by itself guarantee an accurate recovery of the underlying functional regression coefficient. 

\item \emph{Computing time.} The computing time in seconds of each method is also recorded.
\end{itemize}

We point out that the integrals in \eqref{eq:imse-response} and \eqref{eq:imse-beta} cannot be computed exactly, so they are approximated numerically using Simpson's rule. All results are averaged over the 200 simulation runs, and both the mean and the standard deviation are reported. Lower values are better for all these metrics.

\paragraph{Computational environment.}
All the experiments regarding FunSCOPE and FunSCOPE-U have been conducted on a MacBook Pro with an Apple M2 Pro chip (10-core CPU) and 16 GB of RAM, using Python 3.12 \citep{python3}, the \texttt{cpsplines} library \citep{Navarro2023AMC}, and the \texttt{MOSEK 11.0.13} solver \citep{mosek} through its Fusion API for Python. The code is publicly available at \url{https://github.com/mmolerous/FunSCOPE}. For \citet{ghosal2023shape}, we use the code provided by the authors in their supplementary material. It is written in R and has been run on the same machine.

\subsection{Results for SOF regression}
\label{Results SOFR}

We generate data from the SOF regression model following the scenario proposed by \citet{ghosal2023shape}. The observations of the functional covariate \( X(t) \), for $t\in [0,1]$, are given by

\begin{equation*}
x_i(t) = \sum_{j=1}^{20} \psi_{ij} \phi_j(t),
\end{equation*}
where \( \phi_j(t) \) are orthogonal basis polynomials of degree \( j - 1 \), while \( \psi_{ij} \) are independent normal values with mean 0 and variance \( \sigma^2_j = (20 - j + 1) \).

The functional regression coefficient $\beta(t)$ corresponds to the density function of a normal distribution with mean $\mu = 0.5$ and variance $\sigma^2 = 0.01$, thus

\begin{equation}
\label{eq:sof-beta-normal}
\beta(t) =
\frac{1}{\sqrt{2\pi \cdot 0.01}}
\exp \left( - \frac{(t - 0.5)^2}{0.02} \right),
\quad t \in [0,1].
\end{equation}

Observe that $\beta(t)$ in \eqref{eq:sof-beta-normal} is a non-negative function that tends to zero when $t$ goes to infinity (resp. minus infinity). Then, a non-negativity constraint ($\beta(t) \geq 0$) is to be considered when running FunSCOPE and the approach of \cite{ghosal2023shape}.

Finally, the observations of the scalar response variable $Y$, $y_i$, for $i = 1, \dots, n$, are generated according to \eqref{eq:sof}, with $\beta_0 = 0.15$ and $\epsilon_i \sim \mathcal{N}(0, 0.05^2)$.

Table~\ref{tab:sofr_scenario_normal}  presents the comparative results between FunSCOPE-U, FunSCOPE and the method of \citet{ghosal2023shape} for this SOF regression scenario. Each block row of the table corresponds to a different sample size, in which the $\mathrm{MSE}(\hat{\mathbf{y}})$, the $\mathrm{IMSE}(\hat{\beta})$, the running time and the percentage of instances (out of 200) that violated the non-negativity constraint are shown for FunSCOPE, the approach in \cite{ghosal2023shape} and FunSCOPE-U. We observe that in terms of $\mathrm{MSE}(\hat{\mathbf{y}})$ and $\mathrm{IMSE}(\hat{\beta})$, both FunSCOPE-U and FunSCOPE substantially outperform the method of \citet{ghosal2023shape} across all sample sizes, achieving  values that are one to two orders of magnitude lower. Regarding computing time, the proposed methods are considerably faster than that of \citet{ghosal2023shape} across all sample sizes, with computation times between 4 and 22 times lower. Finally, we highlight that if an unconstrained approach is used to estimate an SOF regression model for this scenario as FunSCOPE-U, most of the estimated functional regression coefficients violate this shape requirement for all the sample sizes. However, the shape-constrained approaches prevent this behavior.

\begin{figure}[ht!]
    \centering
    \begin{tabular}{c c c !{\color{gray!40}\vrule} c}
        & \multicolumn{2}{c!{\color{gray!40}\vrule}}{Shape-constrained} & Unconstrained \\
        & FunSCOPE & \citet{ghosal2023shape} & FunSCOPE-U \\[0.5em]
        \rotatebox{90}{$n=25$} &
        \includegraphics[scale=0.15]{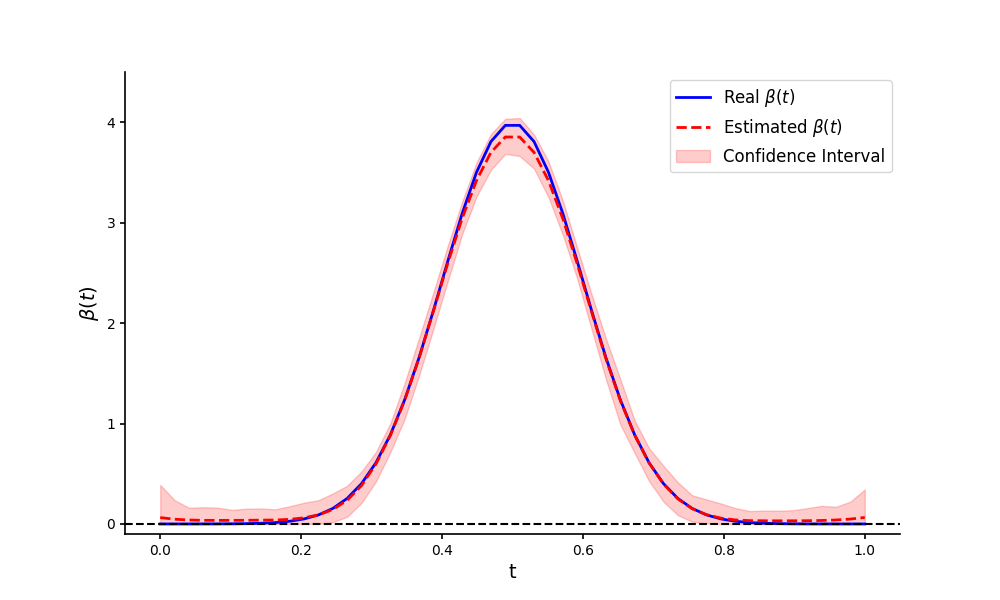} &
        \includegraphics[scale=0.15]{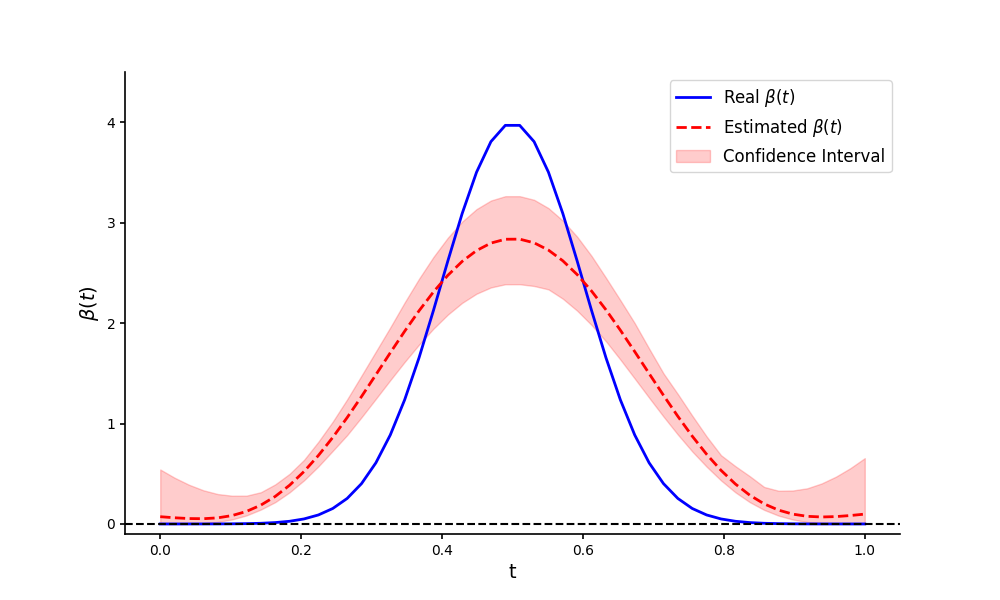} &
        \includegraphics[scale=0.15]{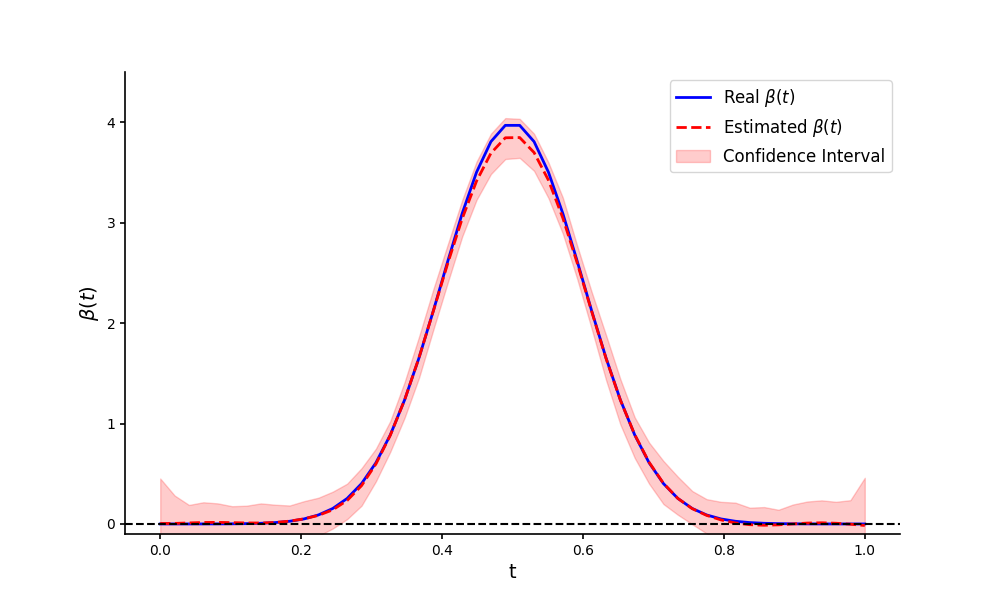} \\[0.5em]
        \rotatebox{90}{$n=50$} &
        \includegraphics[scale=0.15]{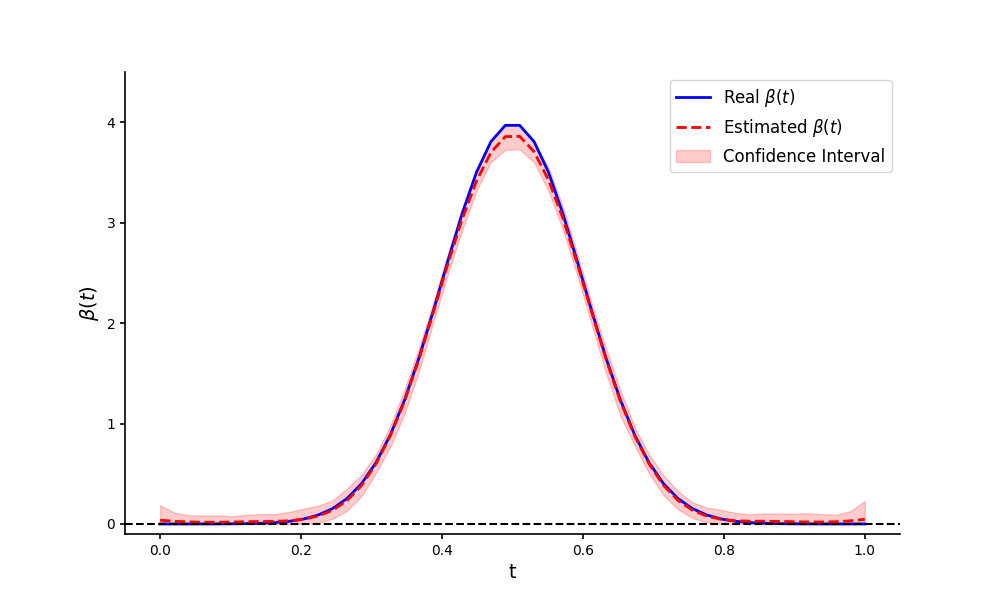} &
        \includegraphics[scale=0.15]{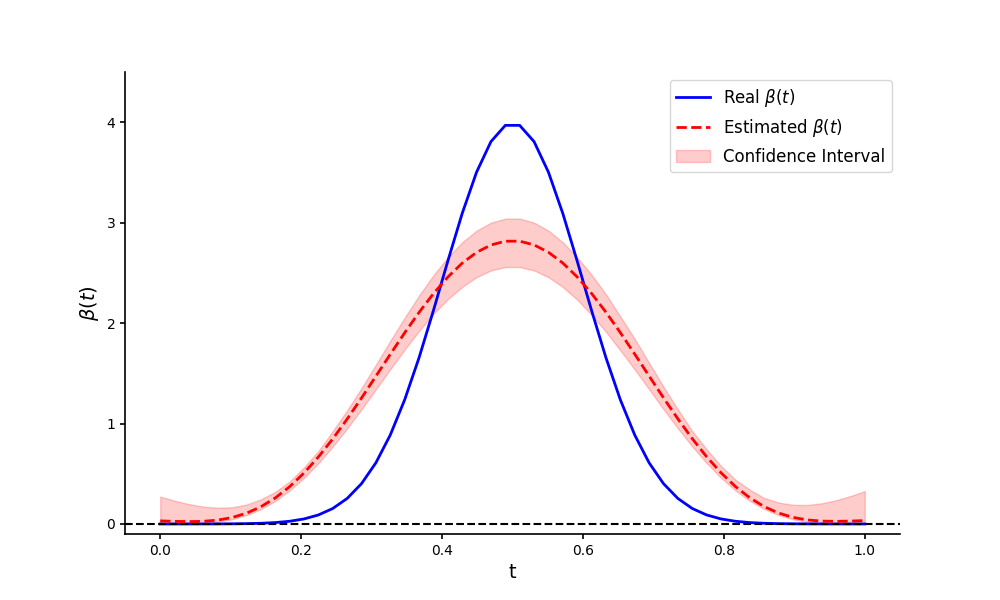} &
        \includegraphics[scale=0.15]{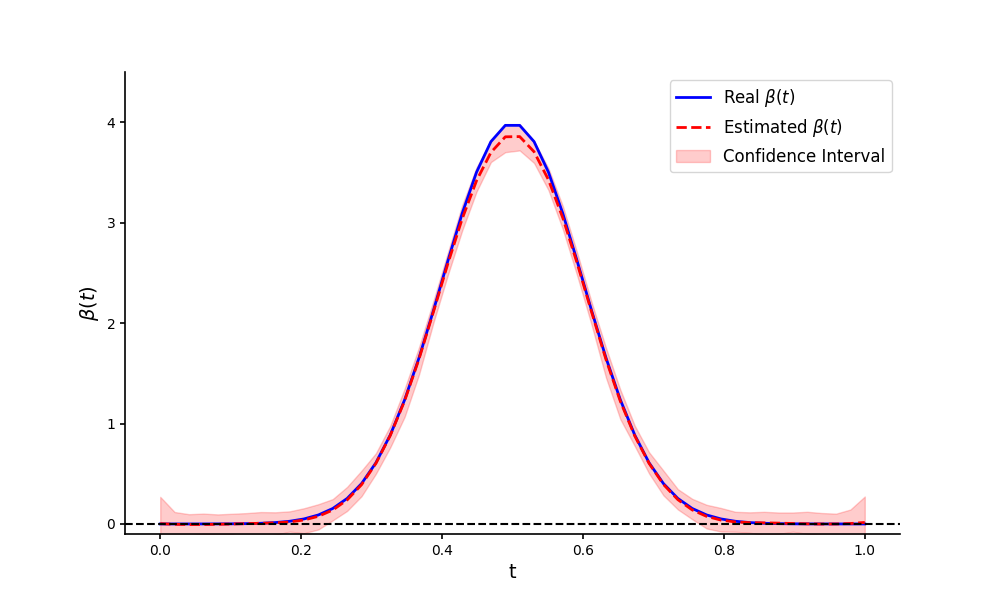} \\[0.5em]
        \rotatebox{90}{$n=100$} &
        \includegraphics[scale=0.15]{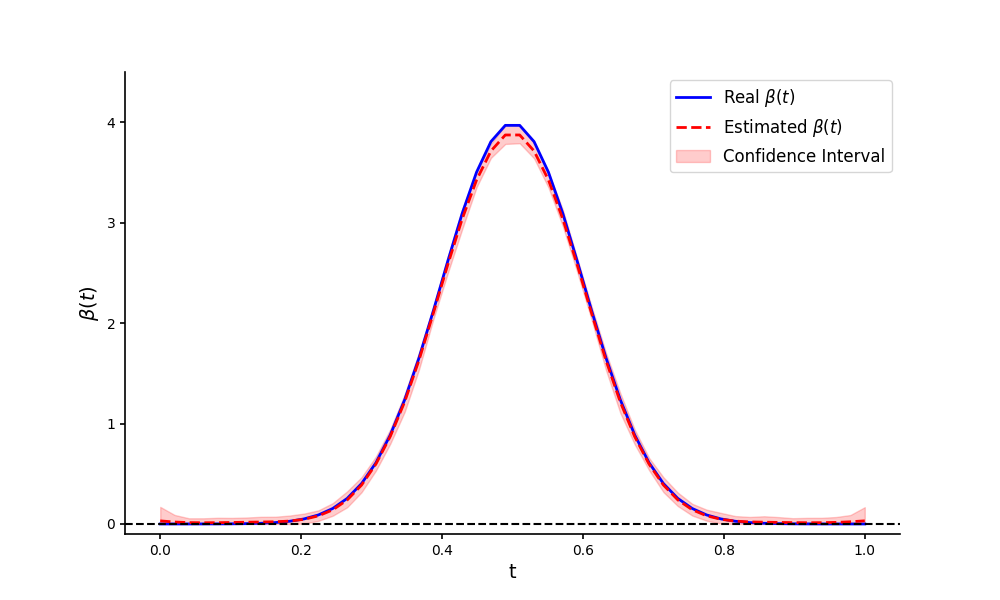} &
        \includegraphics[scale=0.15]{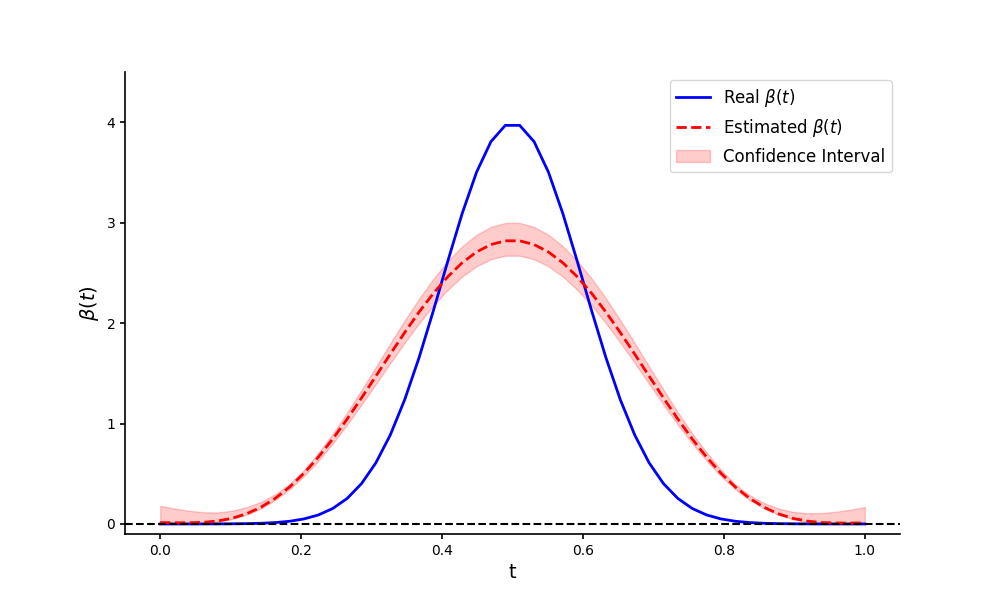} &
        \includegraphics[scale=0.15]{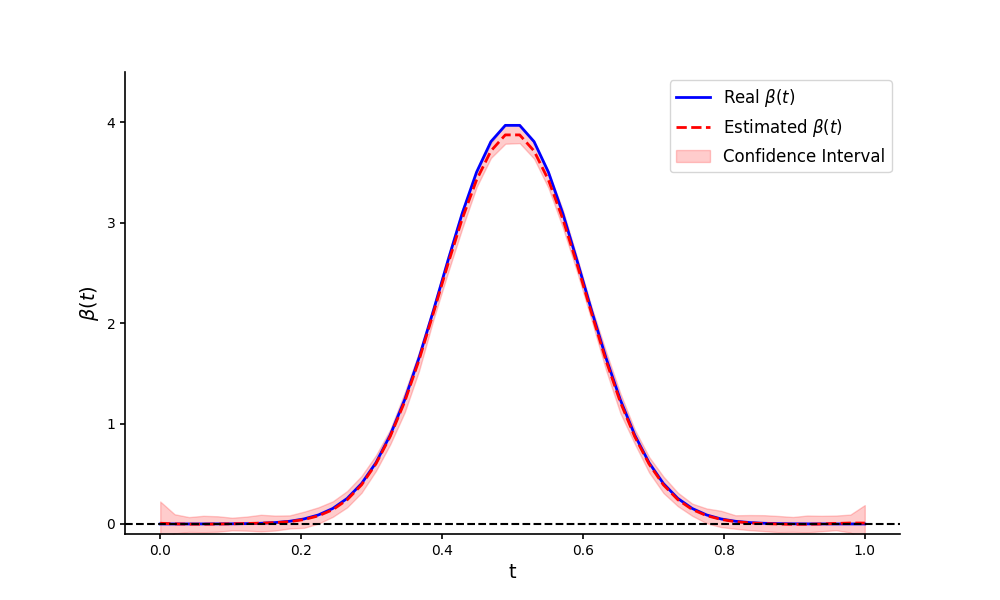} \\
    \end{tabular}
    \caption{SOF regression scenario: true and estimated functional parameter $\beta(t)$ together with 95\% bootstrap confidence intervals of the average estimate for $n \in \{25, 50, 100\}$.}
    \label{fig:Normal_all}
\end{figure}

Figure~\ref{fig:Normal_all} visually confirms the results in Table~\ref{tab:sofr_scenario_normal} across all sample sizes. Each row corresponds to a sample size and each column to the same shape-constrained approaches as in that table. The blue line depicts the true functional regression coefficient in \eqref{eq:sof-beta-normal}, the dashed red line the average of the estimated curves, and the shaded band its 95\% bootstrap confidence interval.  Observe that both FunSCOPE-U and FunSCOPE closely track the true $\beta(t)$ curve and accurately recover its peak. However, a small $\mathrm{IMSE}(\hat{\beta})$ does not guarantee the correct shape: as shown in Table~\ref{tab:sofr_scenario_normal}, FunSCOPE-U violates the non-negativity constraint in most replications, mainly in the tails where $\beta(t)$ is close to zero (see also Figure~\ref{fig:violations}). These violations shrink as $n$ grows, 
but they do not disappear. In contrast, the average estimate of \citet{ghosal2023shape} departs markedly
from the true $\beta(t)$ at all sample sizes: it underestimates the peak and
is flatter and wider than the true curve. Its bands are also wider,
reflecting larger variability. This bias does not decrease with $n$:
$10^{2}\,\mathrm{IMSE}(\hat{\beta})$ only goes from $35.01$ to $32.63$
between $n = 25$ and $n = 100$, while its standard deviation falls from
$3.35$ to $0.52$, which indicates that the error comes mainly from bias
rather than from sampling variability.

\begin{table}[htbp]
\centering
\caption{SOF regression scenario: average (sd) performance across 200 simulations.
Error measures are multiplied by the factor indicated in each row.
}
{\setlength{\aboverulesep}{0pt}\setlength{\belowrulesep}{0pt}\renewcommand{\arraystretch}{1.2}
\begin{tabular}{c|l|cc|c}
\toprule
 & & \multicolumn{2}{c|}{Shape-constrained} & Unconstrained \\
 & Metric & FunSCOPE & \citet{ghosal2023shape} & FunSCOPE-U \\
\midrule
\multirow{4}{*}{$n = 25$}
 & $10^2\,\mathrm{MSE}(\hat{\mathbf{y}})$      & 0.16 (0.08) & 8.26 (2.37)  & 0.12 (0.07) \\
 & $10^2\,\mathrm{IMSE}(\hat{\beta})$ & 0.74 (0.63) & 35.01 (3.35) & 1.23 (0.98) \\
 & Time                               & 0.21s       & 4.59s        & 0.31s       \\
  & Violations             & 0.0\%         & 0.0\%          & 99.5\%        \\
\midrule
\multirow{4}{*}{$n = 50$}
 & $10^2\,\mathrm{MSE}(\hat{\mathbf{y}})$      & 0.22 (0.08) & 8.84 (1.68)  & 0.20 (0.07) \\
 & $10^2\,\mathrm{IMSE}(\hat{\beta})$ & 0.37 (0.19) & 33.14 (1.28) & 0.54 (0.36) \\
 & Time                               & 0.61s       & 4.72s        & 0.60s       \\
  & Violations             & 0.0\%         & 0.0\%          & 100.0\%       \\
\midrule
\multirow{4}{*}{$n = 100$}
 & $10^2\,\mathrm{MSE}(\hat{\mathbf{y}})$      & 0.24 (0.05) & 8.93 (1.17)  & 0.23 (0.05) \\
 & $10^2\,\mathrm{IMSE}(\hat{\beta})$ & 0.24 (0.09) & 32.63 (0.52) & 0.31 (0.11) \\
 & Time                               & 1.21s       & 5.04s        & 1.18s       \\
  & Violations              & 0.0\%         & 0.0\%          & 99.5\%        \\
\bottomrule
\end{tabular}}
\label{tab:sofr_scenario_normal}
\end{table}

\subsection{Results for FOS regression}
\label{Results FOSR}

We generate data from the FOS regression model as follows. Observations of the scalar covariate $X$ are generated assuming $X \sim \text{Unif}(0.5,1.5)$. The functional regression coefficient $\beta(t)$ is defined  for $t \in [0.1, 2.1]$ as the density function of the Weibull distribution with shape parameter $a=0.75$ and scale parameter $b=0.25$, yielding
\begin{equation}
\label{eq:FOS-beta-Weibull}
\beta(t) = 
3\,(4t)^{-1/4}\exp\left(-(4t)^{3/4}\right).
\end{equation}
Because the Weibull density with $a<1$ diverges as $t \to 0$, the domain is restricted to $[0.1, 2.1]$ rather than $[0,1]$ as in the previous scenario. 
Constraints on non-negativity ($\beta(t) \geq 0$) and convexity ($\beta''(t) \geq 0$) are considered as shape-constraints in this scenario.

The functional response $Y(t)$ is then generated according to \eqref{eq:fos}, where the intercept function is set to $\beta_0(t) = 0$ for all $t$, and the error process $\epsilon_i(t)$ is generated as
\begin{equation}
\label{eq:FOS-error-process}
\epsilon_i(t) = c_i \cos(t) + d_i \sin(t) + w_i(t),
\end{equation}
where $c_i, d_i \sim_{iid} \mathcal{N}(0, 0.1^2)$ introduce a smooth, subject-specific correlated component, and $w_i(t) \sim \mathcal{N}(0, 0.1^2)$ is an independent white noise term superimposed at each time point.

\begin{table}[htbp]
\centering
\caption{FOS regression scenario: average (sd) performance across 200 simulations.
Error measures are multiplied by the factor indicated in each row.
}
{\setlength{\aboverulesep}{0pt}\setlength{\belowrulesep}{0pt}\renewcommand{\arraystretch}{1.2}
\begin{tabular}{c|l|c|c}
\toprule
 & & Shape-constrained & Unconstrained \\
 & Metric & FunSCOPE & FunSCOPE-U \\
\midrule
\multirow{4}{*}{$n = 25$}
 & $10^2\,\mathrm{IMSE}(\hat{y})$     & 1.04 (0.22) & 1.03 (0.22) \\
 & $10^2\,\mathrm{IMSE}(\hat{\beta})$ & 0.76 (0.42) & 0.86 (0.49) \\
 & Time                               & 2.63s       & 2.58s       \\
  & Violations             & 0.0\%         & 97.5\%        \\
\midrule
\multirow{4}{*}{$n = 50$}
 & $10^2\,\mathrm{IMSE}(\hat{y})$     & 1.09 (0.16) & 1.09 (0.16) \\
 & $10^2\,\mathrm{IMSE}(\hat{\beta})$ & 0.48 (0.23) & 0.54 (0.29) \\
 & Time                               & 2.77s       & 2.78s       \\
  & Violations            & 0.0\%         & 94.0\%        \\
\midrule
\multirow{4}{*}{$n = 100$}
 & $10^2\,\mathrm{IMSE}(\hat{y})$     & 1.09 (0.11) & 1.09 (0.11) \\
 & $10^2\,\mathrm{IMSE}(\hat{\beta})$ & 0.30 (0.11) & 0.32 (0.12) \\
 & Time                               & 3.29s       & 3.26s       \\
  & Violations   & 0.0\%         & 95.0\%        \\
\bottomrule
\end{tabular}}
\label{tab:fosr_weibull_scenario}
\end{table}

\begin{figure}[ht!]
    \centering
    \begin{tabular}{c c !{\color{gray!40}\vrule} c}
        & Shape-constrained & Unconstrained \\
        & FunSCOPE & FunSCOPE-U \\[0.5em]
        \rotatebox{90}{$n=25$} &
        \includegraphics[scale=0.15]{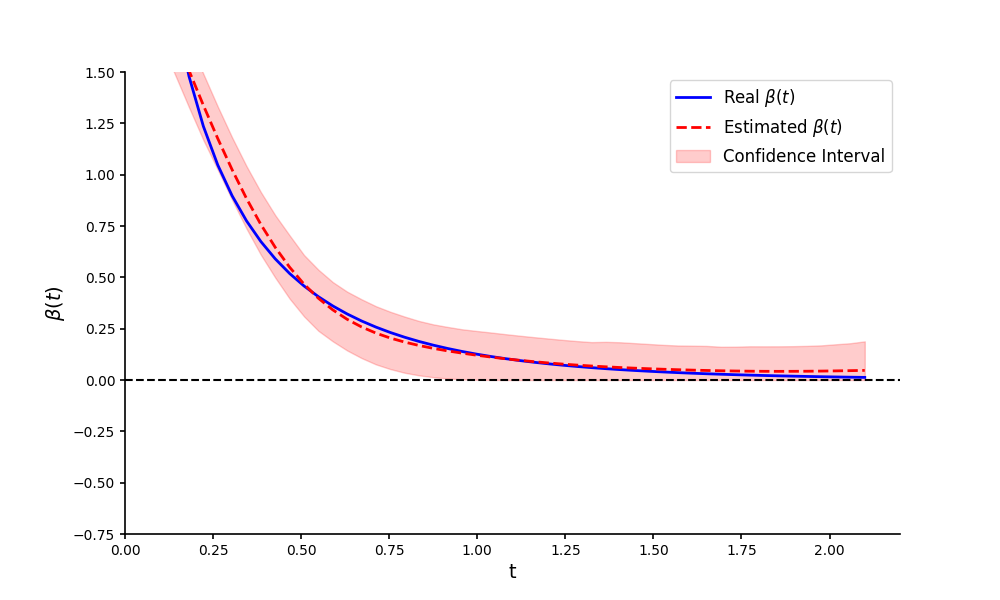} &
        \includegraphics[scale=0.15]{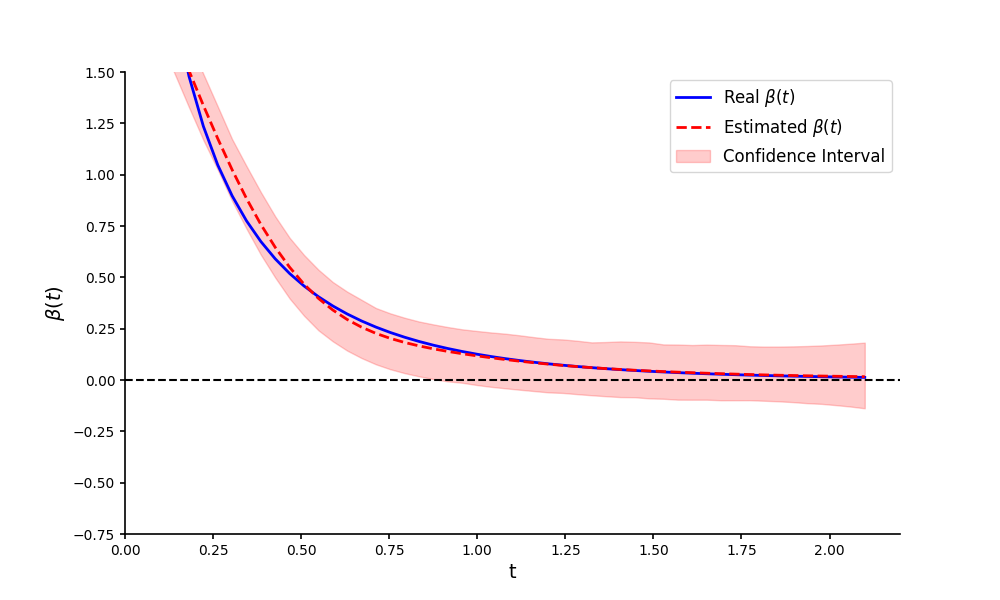} \\[0.5em]
        \rotatebox{90}{$n=50$} &
        \includegraphics[scale=0.15]{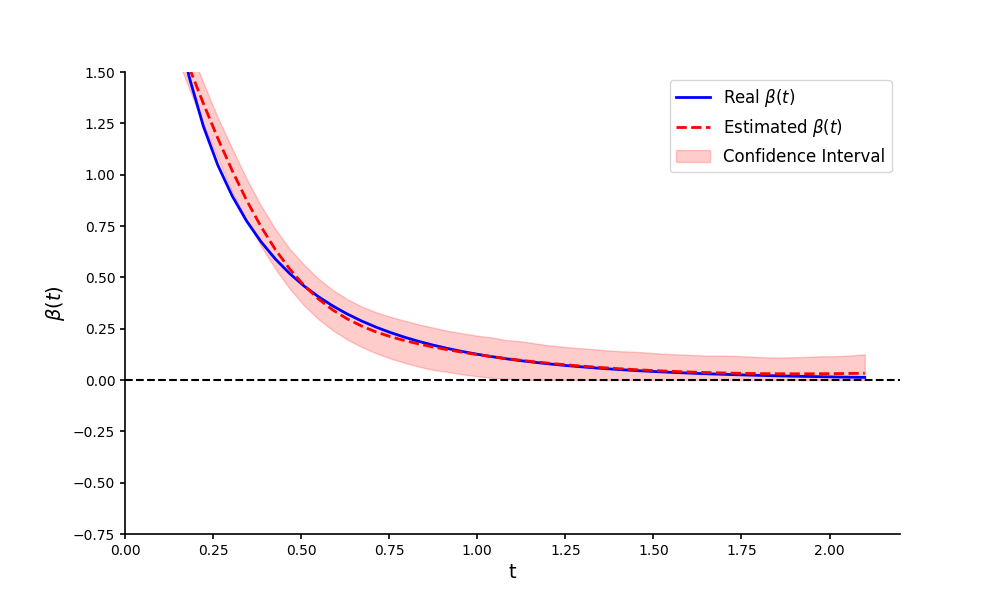} &
        \includegraphics[scale=0.15]{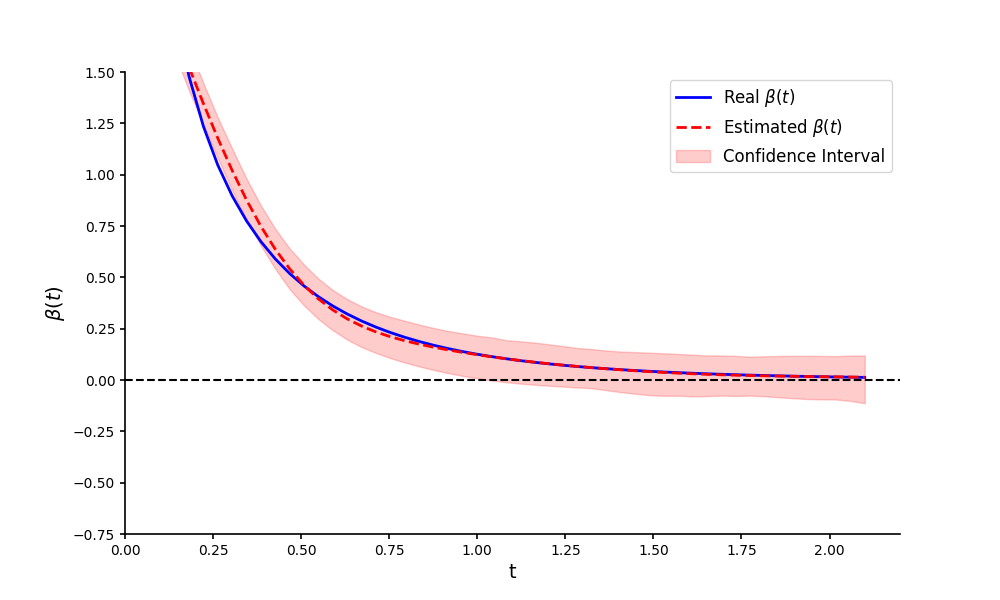} \\[0.5em]
        \rotatebox{90}{$n=100$} &
        \includegraphics[scale=0.15]{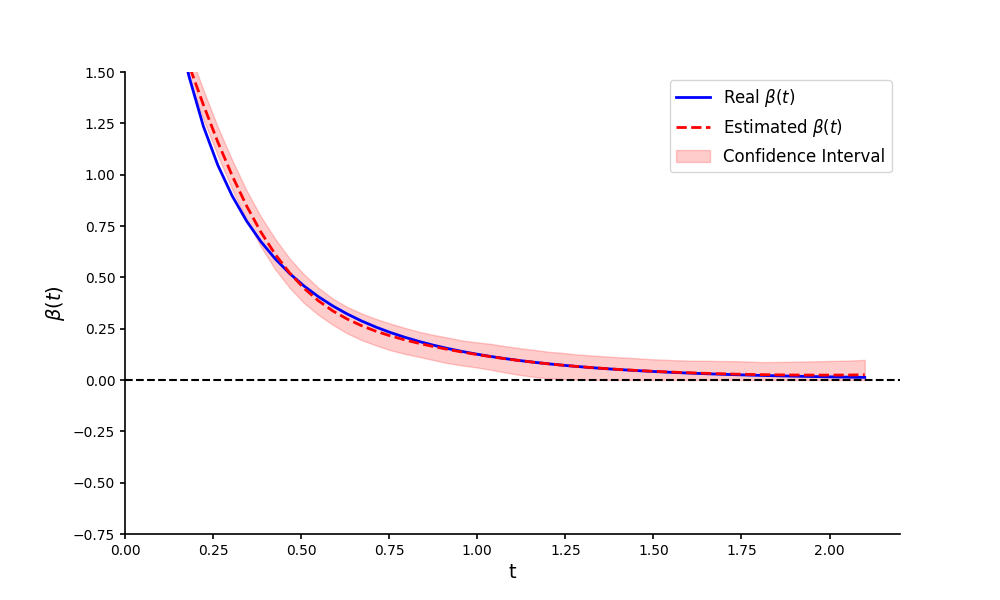} &
        \includegraphics[scale=0.15]{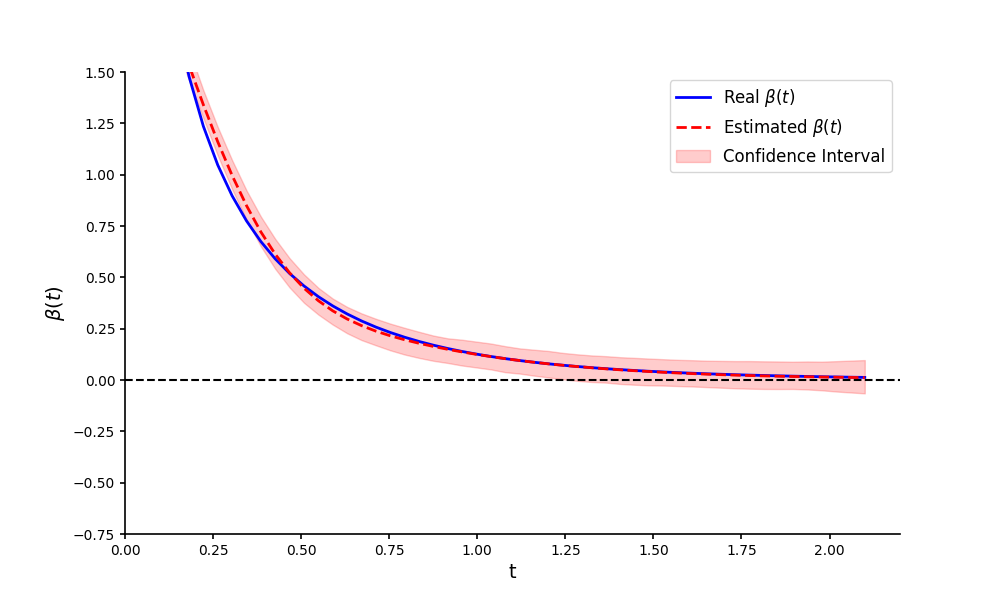} \\
    \end{tabular}
    \caption{FOS regression scenario: estimation of the functional coefficient $\beta(t)$ together with 95\% bootstrap confidence intervals of the average estimate for  $n \in \{25, 50, 100\}$.}
    \label{fig:fosr_weibull_all}
\end{figure}

Table~\ref{tab:fosr_weibull_scenario} and Figure~\ref{fig:fosr_weibull_all} present the comparative results between FunSCOPE-U and FunSCOPE for this FOS scenario. We point out that, as the open-source implementation of \citet{ghosal2023shape} does not support FOS regression, and in that of \citet{han2025inference} the shape constraints considered here, namely non-negativity and convexity, cannot be imposed in a straightforward way, we restrict comparisons to our constrained and unconstrained approaches.

Imposing the non-negativity and convexity constraints (FunSCOPE) consistently improves the recovery of the functional coefficient relative to the unconstrained variant (FunSCOPE-U) across all sample sizes, with the average $\mathrm{IMSE}(\hat{\beta})$ decreasing by approximately 6--12\%, with the largest gains for smaller sample sizes. This advantage is present even for $n=100$, suggesting the constraints continue to provide useful information beyond what the data alone can convey for this scenario. In contrast, $\mathrm{IMSE}(\hat{y})$ remains essentially identical between the two variants, indicating that the shape constraints primarily improve estimation accuracy of $\beta(t)$ without noticeably affecting the fit to the response. As $n$ increases, both variants improve steadily, and computing times remain comparable between the two variants, differing by at most 0.05s on average. Observe again that, in this scenario, the percentage of instances violating the shape constraints remains very high for the unconstrained approach.

Figure~\ref{fig:fosr_weibull_all} confirms this pattern: both variants closely track the true Weibull-shaped $\beta(t)$, with visibly overlapping confidence bands at all three sample sizes. However, the FunSCOPE-U estimates frequently have the wrong shape: at least one of the two constraints is violated in at least 94\% of the replications (Table~\ref{tab:fosr_weibull_scenario}). Non-convexity is the most frequent violation, occurring in about 91\% of them, while negative values, mostly in the right part of the domain where $\beta(t)$ approaches zero, appear in about half of them (see Figure~\ref{fig:violations}). Thus, although the gain in $\mathrm{IMSE}(\hat{\beta})$ is moderate in this scenario, FunSCOPE guarantees by construction that the estimated functional regression coefficient is non-negative and convex over the entire domain. The estimation is most challenging near $t=0.1$, where $\beta(t)$  changes most rapidly, and both methods show wider confidence intervals in that region; as $n$ increases, these bands narrow accordingly.

\subsection{Results for concurrent FOF regression}
\label{Results FOFR}

We generate data from the concurrent FOF regression model \eqref{eq:fofconcurrent} following the scenario proposed by \citet{ghosal2023shape}. The functional covariate $X(t)$ is generated as $x_i(t) = \sum_{j=1}^{5} \psi_{ij} \phi_j(t)$, where $\phi_j(t)$ are orthogonal polynomials of degree $j-1$ and $\psi_{ij}$ are normal scores standardized to zero mean and unit variance, for $i=1,\ldots,n$.

The functional regression coefficient $\beta(t)$ is defined using the hyperbolic tangent function,
\begin{equation}
\label{eq:concurrentFOF-beta-tanh}
\beta(t) = \tanh\left(a(t - t_0)\right), \quad t \in [0,1],
\end{equation}
with $a = 10$ and $t_0 = 0.5$, producing a smooth transition between $-1$ and $1$ centered at $t=0.5$. A monotonicity constraint ($\beta'(t) \geq 0$) is to be considered for the shape-constrained approaches. The intercept function is set to $\beta_0(t) = 0$ for all~$t$ and the error process $\epsilon_i(t)$ is generated as in \eqref{eq:FOS-error-process}. Finally, the functional response variable $Y(t)$ is generated according to  \eqref{eq:fofconcurrent} for $t\in [0,1]$, following the setup described in Section~\ref{Setup}. 

\begin{table}[htbp]
\centering
\caption{Concurrent FOF regression scenario: average (sd) performance across 200 simulations.
Error measures are multiplied by the factor indicated in each row.
}
{\setlength{\aboverulesep}{0pt}\setlength{\belowrulesep}{0pt}\renewcommand{\arraystretch}{1.2}
\begin{tabular}{c|l|cc|c}
\toprule
 & & \multicolumn{2}{c|}{Shape-constrained} & Unconstrained \\
 & Metric & FunSCOPE & \citet{ghosal2023shape} & FunSCOPE-U \\
\midrule
\multirow{4}{*}{$n = 25$}
 & $10^2\,\mathrm{IMSE}(\hat{y})$     & 1.05 (0.25) & 2.27 (0.33) & 1.04 (0.25) \\
 & $10^2\,\mathrm{IMSE}(\hat{\beta})$ & 0.55 (0.41) & 3.48 (1.65) & 0.72 (0.45) \\
 & Time                               & 10.34s      & 33.70s      & 10.23s      \\
  & Violations             & 0.0\%        & 0.0\%        & 100.0\%      \\
\midrule
\multirow{4}{*}{$n = 50$}
 & $10^2\,\mathrm{IMSE}(\hat{y})$     & 1.15 (0.20) & 2.11 (0.19) & 1.14 (0.20) \\
 & $10^2\,\mathrm{IMSE}(\hat{\beta})$ & 0.32 (0.23) & 2.71 (0.04) & 0.38 (0.23) \\
 & Time                               & 10.81s      & 28.11s      & 10.76s      \\
  & Violations             & 0.0\%        & 0.0\%         & 100.0\%      \\
\midrule
\multirow{4}{*}{$n = 100$}
 & $10^2\,\mathrm{IMSE}(\hat{y})$     & 1.14 (0.14) & 2.07 (0.09) & 1.14 (0.14) \\
 & $10^2\,\mathrm{IMSE}(\hat{\beta})$ & 0.19 (0.12) & 2.69 (0.04) & 0.21 (0.12) \\
 & Time                               & 11.86s      & 45.52s      & 11.83s      \\
  & Violations             & 0.0\%         & 0.0\%        & 100.0\%       \\
\bottomrule
\end{tabular}}
\label{tab:fof_tanh_scenario}
\end{table}

\begin{figure}[ht!]
    \centering
    \begin{tabular}{c c c !{\color{gray!40}\vrule} c}
        & \multicolumn{2}{c!{\color{gray!40}\vrule}}{Shape-constrained} & Unconstrained \\
        & FunSCOPE & \citet{ghosal2023shape} & FunSCOPE-U \\[0.5em]
        \rotatebox{90}{$n=25$} &
        \includegraphics[scale=0.15]{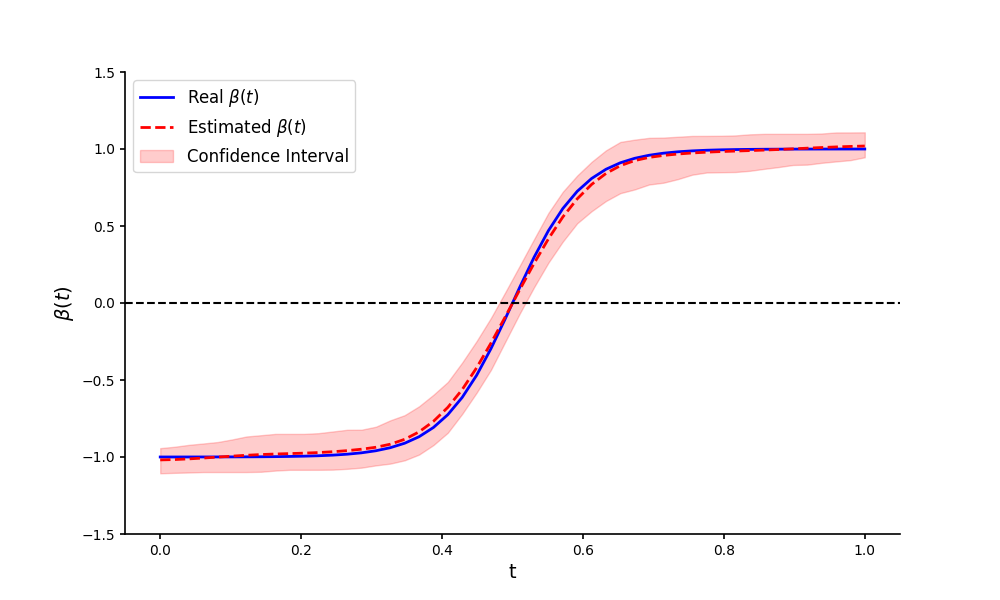} &
        \includegraphics[scale=0.15]{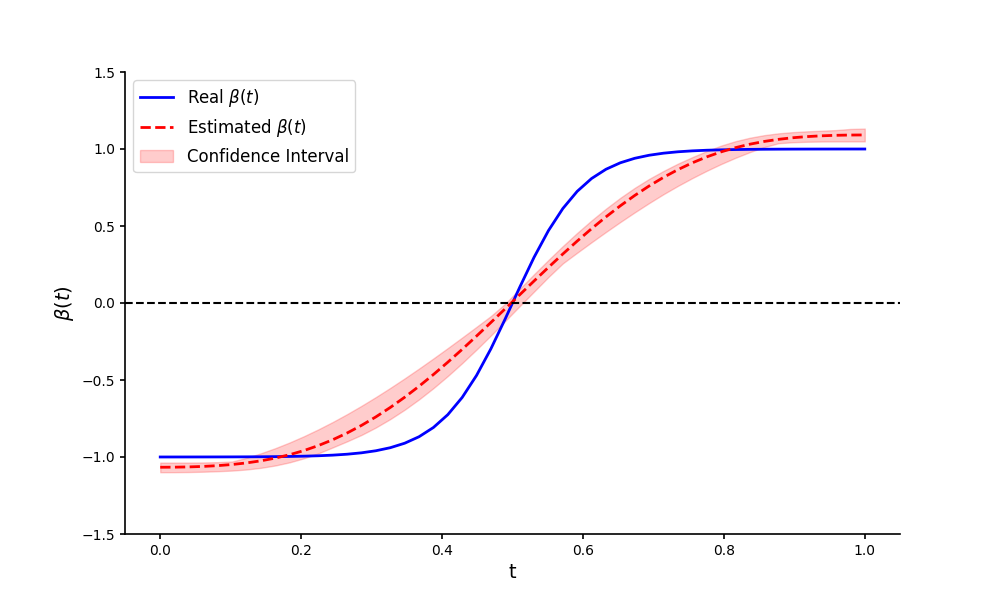} &
        \includegraphics[scale=0.15]{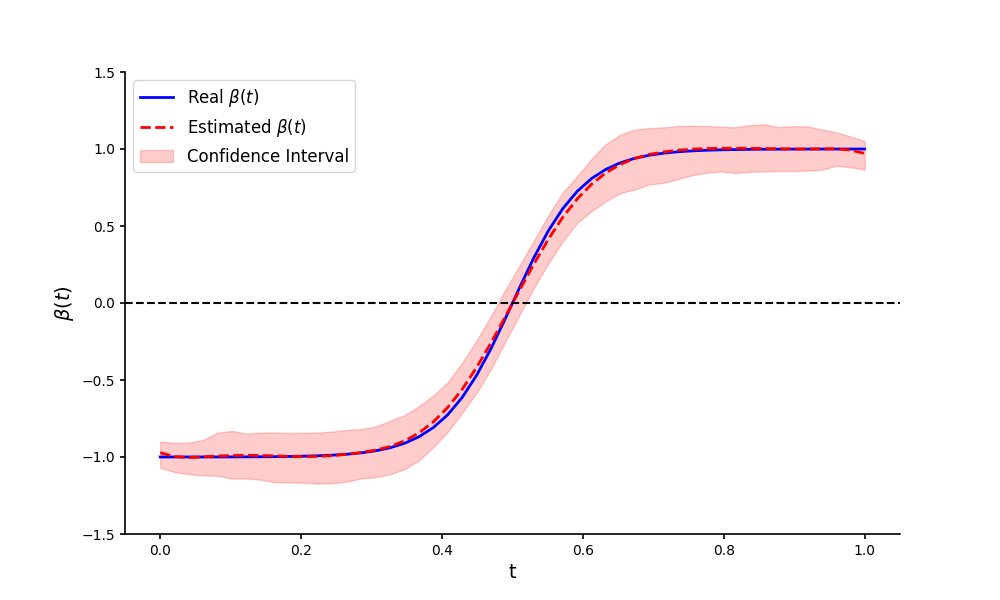} \\[0.5em]
        \rotatebox{90}{$n=50$} &
        \includegraphics[scale=0.15]{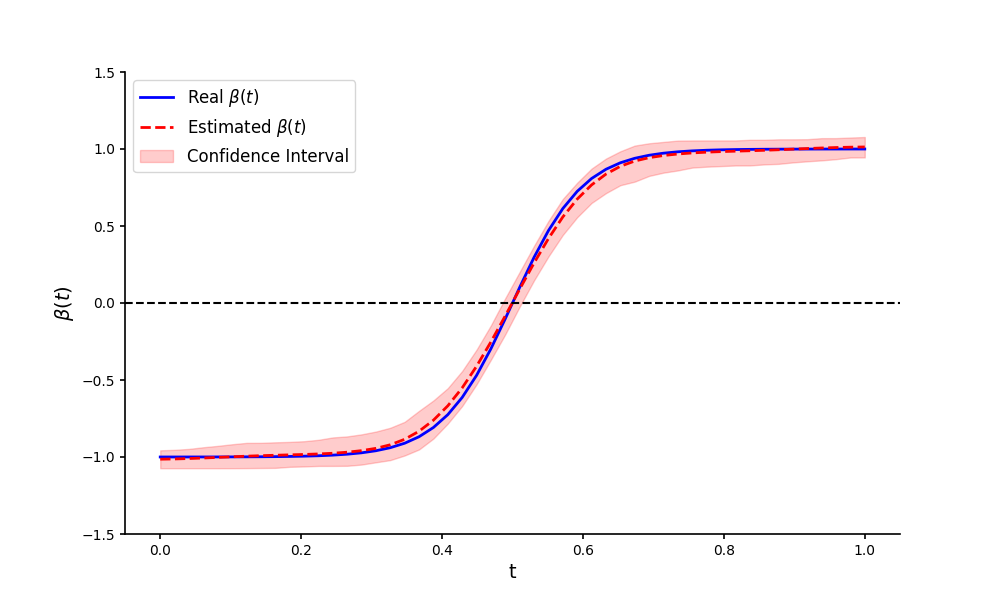} &
        \includegraphics[scale=0.15]{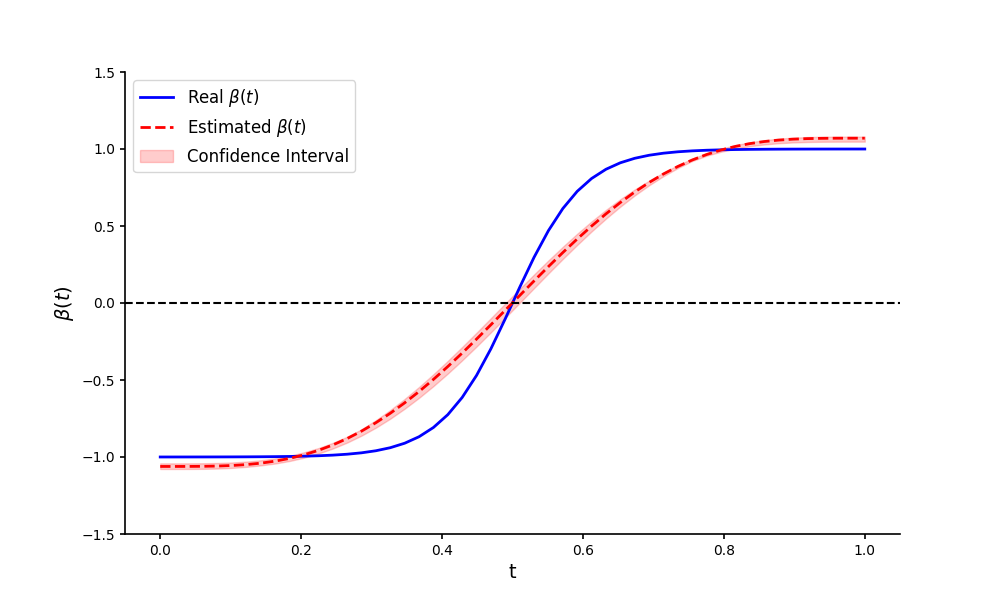} &
        \includegraphics[scale=0.15]{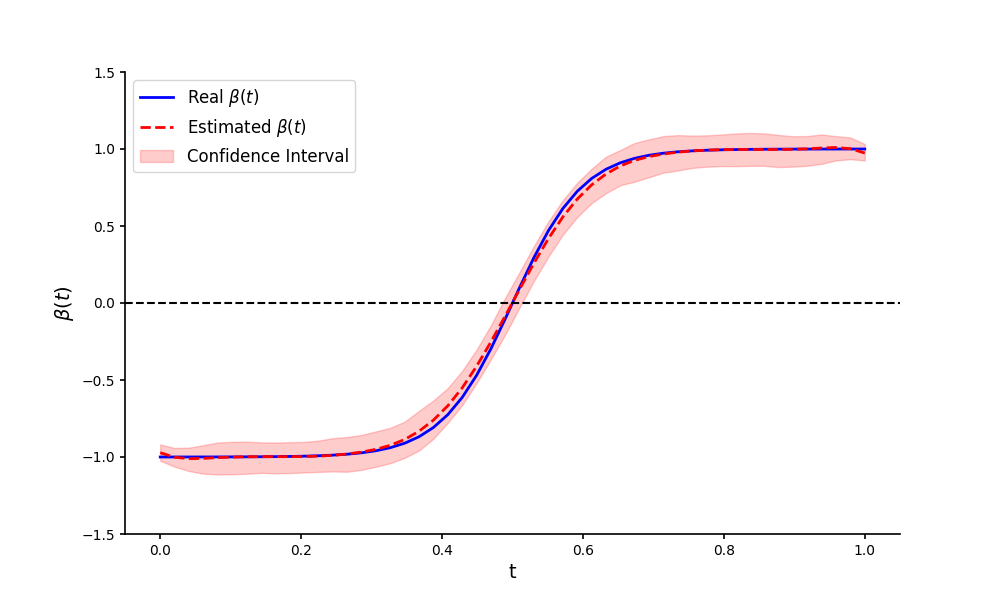} \\[0.5em]
        \rotatebox{90}{$n=100$} &
        \includegraphics[scale=0.15]{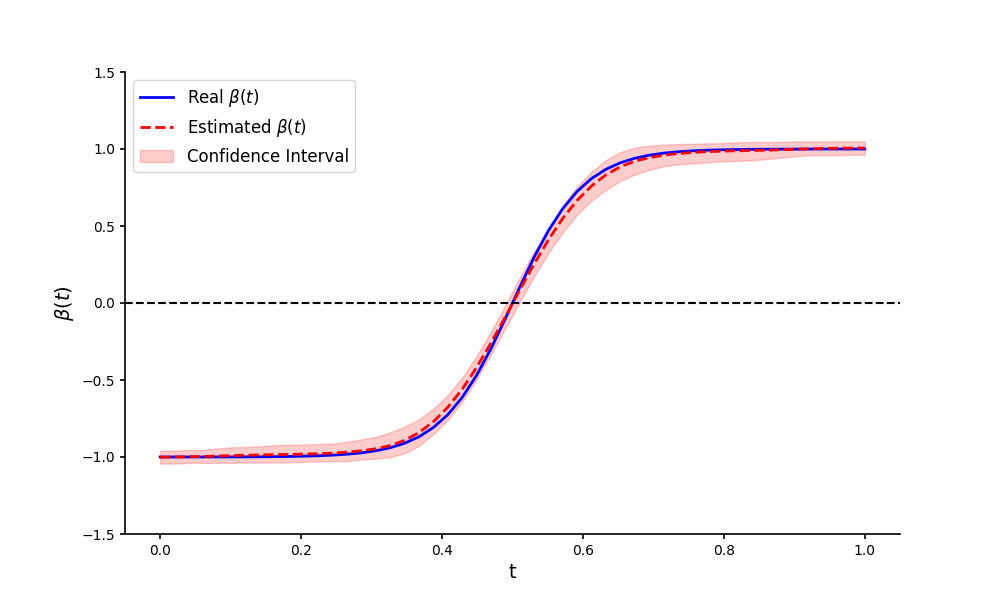} &
        \includegraphics[scale=0.15]{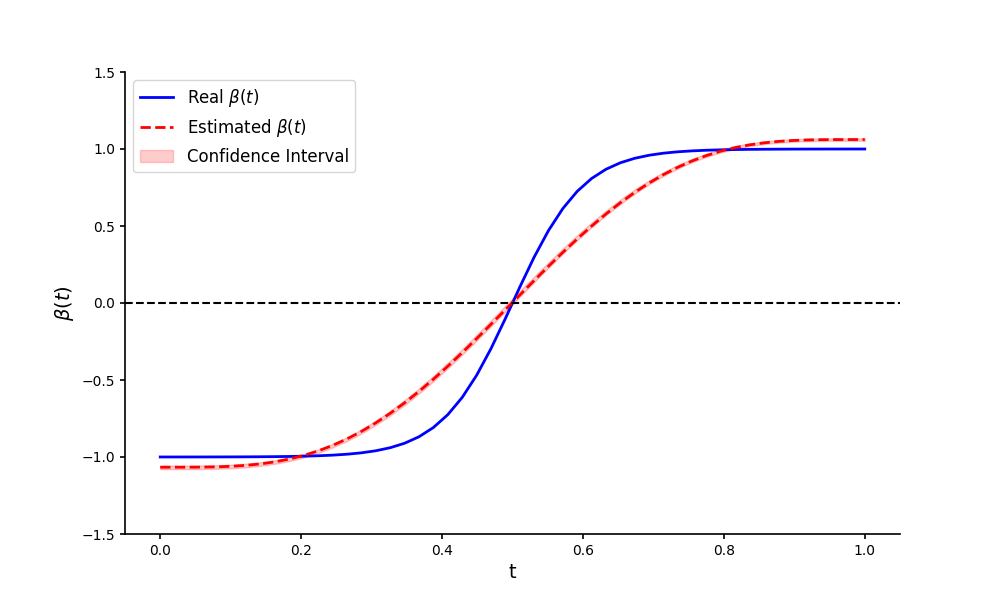} &
        \includegraphics[scale=0.15]{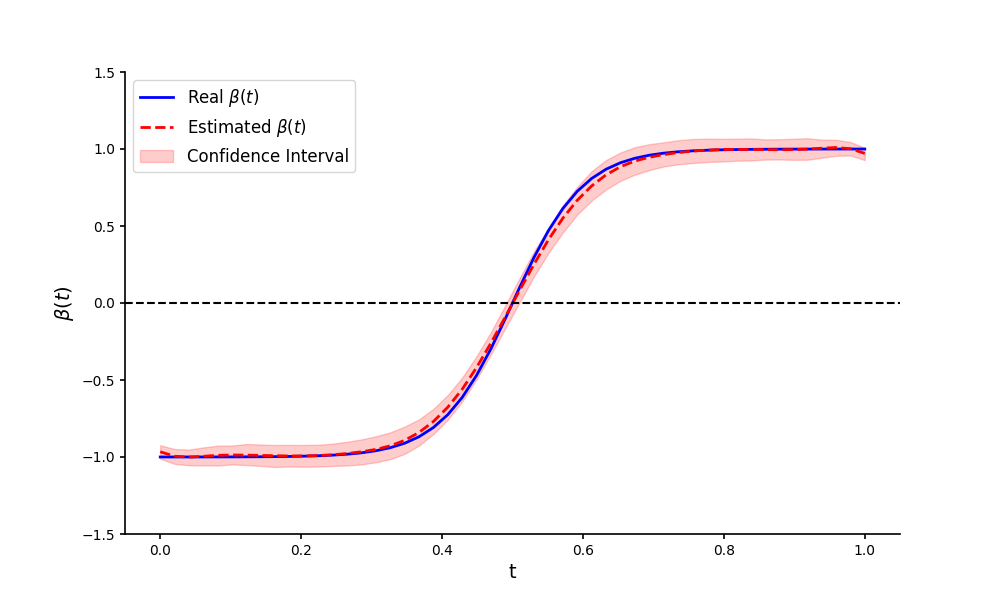} \\
    \end{tabular}
    \caption{Concurrent FOF regression scenario: estimation of the functional coefficient $\beta(t)$ together with 95\% bootstrap confidence intervals of the average estimate for $n \in \{25, 50, 100\}$.}
    \label{fig:tanh_all}
\end{figure}

Table~\ref{tab:fof_tanh_scenario} and Figure~\ref{fig:tanh_all} present the comparative results between FunSCOPE-U, FunSCOPE and the method of \citet{ghosal2023shape} for this concurrent FOF scenario.

In terms of accuracy, both FunSCOPE-U and FunSCOPE substantially outperform the method of \citet{ghosal2023shape} in the recovery of the functional coefficient across all sample sizes, with $\mathrm{IMSE}(\hat{\beta})$ values roughly five to fourteen times lower. As noted in Section~\ref{Setup}, the $\mathrm{IMSE}(\hat{y})$ values are not directly comparable although they are shown for illustrative purposes.

Imposing the monotonicity constraint on the first derivative (FunSCOPE) further improves the recovery of the functional coefficient relative to the unconstrained variant (FunSCOPE-U), with this advantage being most pronounced for the smallest sample size ($n=25$), where the average $\mathrm{IMSE}(\hat{\beta})$ decreases by about 24\% thanks to the prior knowledge that $\beta(t)$ is increasing. As $n$ increases, the performance gap between the two variants narrows, in line with the pattern observed in the previous scenarios. Regarding computing time, the proposed methods remain markedly faster than that of \citet{ghosal2023shape} across all sample sizes, being between 2.6 and 3.9 times lower. Furthermore, observe that the functional regression coefficients estimated using the unconstrained approach violate the shape constraints in all the 200 instances.

Figure~\ref{fig:tanh_all} visually confirms these results. Both FunSCOPE and FunSCOPE-U closely track the true sigmoidal curve, including its sharp transition around $t=0.5$. FunSCOPE-U, however, shows small oscillations near the boundaries $t=0$ and $t=1$, together with a wider and more irregular confidence band, since no shape restriction is imposed on the estimate. As a result, it fails to be increasing in every replication (Table~\ref{tab:fof_tanh_scenario} and Figure~\ref{fig:violations}), and these decreases are not negligible.
FunSCOPE, which enforces the correct monotonicity, removes these oscillations and yields a narrower band. The average estimate of \citet{ghosal2023shape}, in contrast, is clearly biased:
it is a flatter sigmoid that underestimates the steepness of $\beta(t)$
around $t = 0.5$ and slightly overshoots its range near the boundaries.
The true curve lies outside the 95\% confidence band of the average estimate
over a large part of the domain, so this bias cannot be attributed to
simulation noise. The bias
does not decrease with $n$: $10^{2}\,\mathrm{IMSE}(\hat{\beta})$ levels off
at about $2.7$ from $n = 50$ onwards, with a standard deviation of only
$0.04$.

\subsection{Summary}
\label{subsec:summary}

\begin{figure}[ht!]
    \centering
    \includegraphics[trim={0 0 0 0.98cm},clip, width=\textwidth]{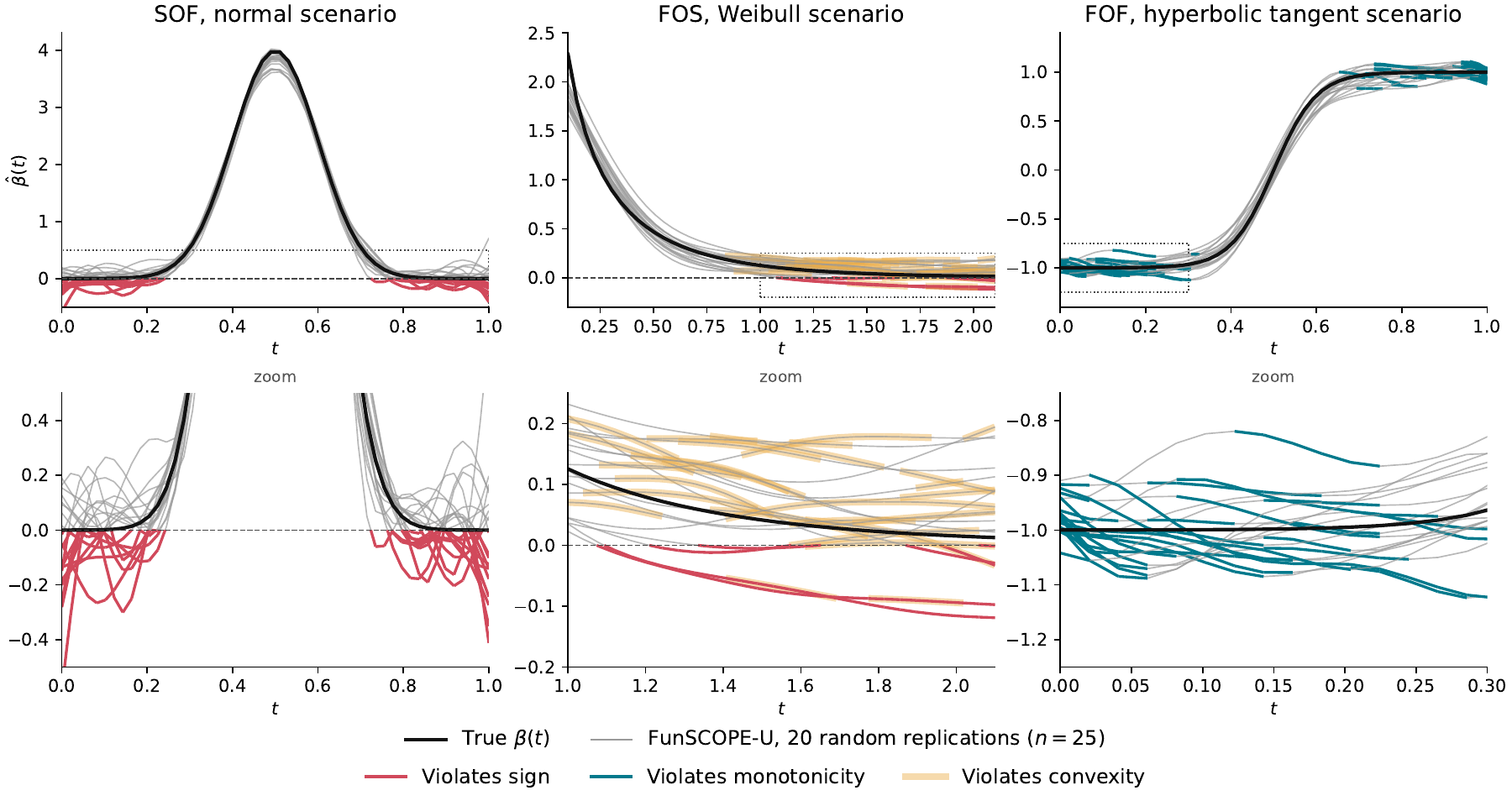}
    \caption{Shape violations of some estimated functional regression coefficients by FunSCOPE-U: 20 randomly selected replications with $n=25$ for the SOF (left), FOS (center) and concurrent FOF (right) regression scenarios. Highlighted segments indicate where the estimates violate the sign (SOF, FOS), monotonicity (FOF) or convexity (FOS) constraints. The bottom row zooms in on the regions marked with dotted rectangles. 
    }
    \label{fig:violations}
\end{figure}

The simulation study across SOF, FOS and concurrent FOF linear regression models has allowed for a controlled comparison of the shape-constrained functional regression approaches FunSCOPE and that by \cite{ghosal2023shape}, jointly with the unconstrained counterpart of the first one, FunSCOPE-U, from which two main conclusions can be drawn.

First, imposing the shape constraints matters even when the unconstrained estimator is already accurate: FunSCOPE-U violates the shape of the true functional regression coefficient in almost every replication of the SOF and FOF scenarios and in at least 94\% of those of the FOS scenario, and although these violations shrink as $n$ increases, they do not disappear, whereas FunSCOPE satisfies the constraints by construction while also reducing $\mathrm{IMSE}(\hat{\beta})$. Figure~\ref{fig:violations} illustrates the shape violations of FunSCOPE-U reported in Tables~\ref{tab:sofr_scenario_normal}, \ref{tab:fosr_weibull_scenario} and~\ref{tab:fof_tanh_scenario} for 20 randomly selected replications with $n=25$ in each scenario, highlighting the parts of the estimates where the constraints are violated.

Second, FunSCOPE outperforms the state-of-the-art approach of \citet{ghosal2023shape}, as implemented by its authors, in both accuracy and computing time.
The bias of the latter does not decrease as $n$ grows, so its remaining
error is not due to a lack of data. A plausible explanation lies in its
shape constraints, which are imposed through conditions on the Bernstein
coefficients that are sufficient but not necessary and may therefore exclude
good approximations of $\beta(t)$ from the feasible set.

\section{Conclusions and perspectives}
\label{Conclusions and perspectives}

In this paper, we have proposed FunSCOPE, a unified framework for shape-constrained functional regression using P-splines that covers scalar-on-function, function-on-scalar, and function-on-function models, and in which any combination of sign, monotonicity, and curvature constraints can be imposed on the functional regression coefficient. Building on a basis representation of the functional data, the shapes are expressed as conditions on the spline coefficients, which leads to a conic optimization problem. For univariate coefficients, as in the scalar-on-function, function-on-scalar, and concurrent function-on-function models, these conditions are both necessary and sufficient, in contrast to competing approaches, which only provide sufficient conditions. 
For bivariate coefficients, as in general function-on-function regression, they are neither necessary nor sufficient, although the smoothness induced by the P-spline penalty makes violations of the shape constraints unlikely in practice, as observed by \citet{navarro2024} in shape-constrained multivariate smooth regression. Our numerical experiments show that FunSCOPE outperforms the competing method of \citet{ghosal2023shape} in terms of estimation error and computational time.

Several extensions of FunSCOPE are of interest. First, moving beyond the Gaussian setting to functional generalized linear models would broaden the applicability of FunSCOPE to binary, count, or other non-Gaussian responses. Second, extending FunSCOPE to nonlinear functional regression models, such as functional additive models, would allow shape constraints to be imposed on the nonlinear effects of the functional predictor. Finally, FunSCOPE could be extended to multidimensional functional data, such as images or surfaces, where each observation depends on more than one argument. Tensor-product P-splines provide a natural basis in this setting, although formulating the constraints over a multidimensional domain poses additional challenges.

\section*{Acknowledgments}

This research is supported by the I+D+i projects PID2022-137243OB-I00 and PID2022-137818OB-I00, and the IMUS-María de Maeztu grant CEX2024-001517-M – Apoyo a Unidades de Excelencia María de Maeztu, funded by MICIU/AEI/10.13039/501100011033, the first two also by FEDER/UE; the grant DGP\_POST\_2024\_00630, funded by JA/CUII and FSE+; the research project P2025-0040\_PRMO, funded by the FMJH Program PGMO; and the L'Oréal-UNESCO "For Women in Science" 2023 Award. This support is gratefully acknowledged. This work was initiated through the JDC2023-050674-I grant, funded by FSE+ and MICIU/AEI/10.13039/501100011033, as a continuation of Jesús Secilla's Bachelor's Thesis, which laid the groundwork for this work, and to whom we are also grateful. 

\bibliographystyle{plainnat} 
\bibliography{FunSCOPE}

\newpage
\appendix

\section{Estimation of the FOS linear regression model \texorpdfstring{\eqref{eq:fos-basis}}{}
         \label{app:FOS-errorsdep}}

In this appendix, we detail the calculations leading to the expressions of $\mathbf{u}$ and $\mathbf{Z}$ in \eqref{eq:uZfos} and \eqref{eq:uZfos-gls} for the FOS linear regression model \eqref{eq:fos-basis}, under uncorrelated and correlated errors, respectively.

\subsection{The IMSE under uncorrelated errors \label{subsec:app-fos-uncorrelated}}

Recall that $\mathbf{C}\in\mathbb{R}^{n\times{(k+d)}}$ is the matrix that has $\mathbf{c}_i^\top$ in its $i-$th row, $i=1,\ldots,n,$ $\boldsymbol{\Psi} = \int_0^T \mathbf{B}(t)^\top \mathbf{B}(t)\,dt,$ $\mathbf{x} = (x_1,\ldots,x_n)^\top,$ $\vect(\cdot)$ is the vectorization operator and $\otimes$ the Kronecker product. Furthermore, let $\Tr(\cdot)$ be the trace operator. Because the response is functional, the least-squares criterion corresponds to minimizing an integrated mean squared error (IMSE) over the domain $[0,T]$, rather than the standard mean squared error used for scalar responses. This IMSE can be written as follows:
\begingroup
\allowdisplaybreaks
\begin{align}
    & \displaystyle\int_{0}^T\left\|\mathbf{B}(t)\mathbf{C}^\top - \mathbf{B}(t)\mathbf{b}\mathbf{x}^\top\right\|_2^2 dt \nonumber \\
    & =
    \displaystyle\int_{0}^T\left(\mathbf{B}(t)\mathbf{C}^\top - \mathbf{B}(t)\mathbf{b}\mathbf{x}^\top\right)\left(\mathbf{B}(t)\mathbf{C}^\top - \mathbf{B}(t)\mathbf{b}\mathbf{x}^\top\right)^\top dt \nonumber  \\
    & = \displaystyle\int_{0}^T \Tr\left\{ \left(\mathbf{B}(t)\mathbf{C}^\top - \mathbf{B}(t)\mathbf{b}\mathbf{x}^\top\right)\left(\mathbf{B}(t)\mathbf{C}^\top - \mathbf{B}(t)\mathbf{b}\mathbf{x}^\top\right)^\top \right\} dt \nonumber  \\
    & = \displaystyle\int_{0}^T \Tr \left\{\mathbf{B}(t) (\mathbf{C}^\top - \mathbf{b}\mathbf{x}^\top) (\mathbf{C}^\top - \mathbf{b}\mathbf{x}^\top)^\top \mathbf{B}(t)^\top  \right\} dt \nonumber  \\
    & = \displaystyle\int_{0}^T \Tr \left\{ (\mathbf{C}^\top - \mathbf{b}\mathbf{x}^\top) (\mathbf{C}^\top - \mathbf{b}\mathbf{x}^\top)^\top \mathbf{B}(t)^\top \mathbf{B}(t) \right\} dt \nonumber  \\
    & =\Tr \left\{(\mathbf{C}^\top - \mathbf{b}\mathbf{x}^\top) (\mathbf{C}^\top - \mathbf{b}\mathbf{x}^\top)^\top  \boldsymbol{\Psi} \right\}  \nonumber  \\
    & = \Tr  \left\{ \left[ \boldsymbol{\Psi}^{1/2}(\mathbf{C}^\top - \mathbf{b}\mathbf{x}^\top)  \right] \left[ \boldsymbol{\Psi}^{1/2}(\mathbf{C}^\top - \mathbf{b}\mathbf{x}^\top)  \right]^\top \right\}  \nonumber  \\
    & = \left(\vect(\mathbf{C}\boldsymbol{\Psi}^{1/2}) - \vect(\mathbf{x}\mathbf{b}^\top\boldsymbol{\Psi}^{1/2}) \right)^\top \left(\vect(\mathbf{C}\boldsymbol{\Psi}^{1/2}) - \vect(\mathbf{x}\mathbf{b}^\top\boldsymbol{\Psi}^{1/2}) \right) \nonumber  \\
    &= \left(\vect(\mathbf{C}\boldsymbol{\Psi}^{1/2}) - (\boldsymbol{\Psi}^{1/2}\otimes \mathbf{x})\vect(\mathbf{b}^\top) \right)^\top \left(\vect(\mathbf{C}\boldsymbol{\Psi}^{1/2}) - (\boldsymbol{\Psi}^{1/2}\otimes \mathbf{x})\vect(\mathbf{b}^\top) \right) \nonumber  \\
    & = \left\| \vect(\mathbf{C}\boldsymbol{\Psi}^{1/2}) - (\boldsymbol{\Psi}^{1/2}\otimes \mathbf{x})\mathbf{b}  \right\|_2^2, \label{eq:fos_imse_indep}
\end{align}
\endgroup
which is the least-squares term $\|\mathbf{u}-\mathbf{Z}\boldsymbol{\delta}\|_2^2$ of \eqref{eq:frm} with $\mathbf{u}$, $\mathbf{Z}$ and $\boldsymbol{\delta}$ as in \eqref{eq:uZfos}.

\subsection{The GLS criterion under correlated errors \label{subsec:app-fos-correlated}}

Let us consider $\mathbf{E} = \left(\mathbf{C}-\mathbf{x}\mathbf{b}^\top\right)\boldsymbol{\Psi}^{1/2}
\in\mathbb{R}^{n\times(k+d)}$, whose $i$-th row is $\boldsymbol{\rho}_i^\top$ with
\begin{equation}\label{eq:rho}
\boldsymbol{\rho}_i \;=\; \boldsymbol{\Psi}^{1/2}\left(\mathbf{c}_i - x_i\mathbf{b}\right),
\qquad i=1,\ldots,n.
\end{equation}
Vectorizing $\mathbf{E}$ and using $\vect(\mathbf{A}\mathbf{X}\mathbf{B}) = (\mathbf{B}^\top\otimes\mathbf{A})\vect(\mathbf{X})$, together with the symmetry of $\boldsymbol{\Psi}^{1/2}$, we obtain
\begin{equation}\label{eq:vecE}
    \vect(\mathbf{E}) \;=\; \vect(\mathbf{C}\boldsymbol{\Psi}^{1/2}) - (\boldsymbol{\Psi}^{1/2}\otimes\mathbf{x})\mathbf{b},
\end{equation}
which is precisely the residual vector of \eqref{eq:fos_imse_indep}, and the IMSE reads $\sum_{i=1}^n\|\boldsymbol{\rho}_i\|_2^2$. Minimizing that sum of squares is the appropriate criterion only when the entries of $\vect(\mathbf{E})$ are uncorrelated with constant variance; otherwise, the residuals must be weighted by the inverse of their covariance matrix.

Since the rows of $\mathbf{E}$ are independent across individuals with a common covariance matrix $\boldsymbol{\Sigma}=\mathrm{Cov}(\boldsymbol{\rho}_i)\in\mathbb{R}^{(k+d)\times(k+d)}$, it holds that
$\mathrm{Cov}\left(\vect(\mathbf{E})\right)=\boldsymbol{\Sigma}\otimes I_n$, where $I_n$ denotes the
identity matrix of size $n$; the criterion \eqref{eq:fos_imse_indep} is thus the one obtained when
$\boldsymbol{\Sigma}=\sigma^2 I_{k+d}$. In practice, $\boldsymbol{\Sigma}$ is estimated from the
residual curves of the first stage, i.e., under uncorrelated errors,
$\widehat{\boldsymbol{\rho}}_i=\boldsymbol{\Psi}^{1/2}\left(\mathbf{c}_i-x_i\widehat{\mathbf{b}}\right)$,
$i=1,\ldots,n$, by Functional Principal Component Analysis (FPCA) on
$\widehat{\boldsymbol{\varepsilon}}=(\widehat{\boldsymbol{\rho}}_1,\ldots,
\widehat{\boldsymbol{\rho}}_n)^\top\in\mathbb{R}^{n\times(k+d)}$ \citep{yao2005functional}, yielding
$\widehat{\boldsymbol{\Sigma}}$. The generalized version of
the least-squares criterion \eqref{eq:fos_imse_indep} then becomes
\begin{align}
    &  \left(\vect(\mathbf{C}\boldsymbol{\Psi}^{1/2}) - (\boldsymbol{\Psi}^{1/2}\otimes \mathbf{x})\mathbf{b} \right)^\top \left( \widehat{\boldsymbol{\Sigma}} \otimes  I_n\right)^{-1} \left(\vect(\mathbf{C}\boldsymbol{\Psi}^{1/2}) - (\boldsymbol{\Psi}^{1/2}\otimes \mathbf{x})\mathbf{b} \right) \nonumber\\
    & \qquad = \vect(\mathbf{E})^\top \left(\widehat{\boldsymbol{\Sigma}}^{-1/2}\otimes I_n\right)\left(\widehat{\boldsymbol{\Sigma}}^{-1/2}\otimes I_n\right)\vect(\mathbf{E}) \nonumber\\
    & \qquad = \left\|\left(\widehat{\boldsymbol{\Sigma}}^{-1/2}\otimes I_n\right)\vect(\mathbf{E})\right\|_2^2
    \;=\; \left\|\vect\left(\mathbf{E}\,\widehat{\boldsymbol{\Sigma}}^{-1/2}\right)\right\|_2^2 \nonumber\\
    & \qquad = \left\|\vect(\mathbf{C}\boldsymbol{\Psi}^{1/2}\widehat{\boldsymbol{\Sigma}}^{-1/2} )  - (\widehat{\boldsymbol{\Sigma}}^{-1/2}\boldsymbol{\Psi}^{1/2}\otimes \mathbf{x} )\mathbf{b} \right\|_2^2, \label{eq:fos-genleastsquares}
\end{align}
where the first equality follows from \eqref{eq:vecE} and
$(\widehat{\boldsymbol{\Sigma}}\otimes I_n)^{-1} = \widehat{\boldsymbol{\Sigma}}^{-1}\otimes I_n = (\widehat{\boldsymbol{\Sigma}}^{-1/2}\otimes I_n)(\widehat{\boldsymbol{\Sigma}}^{-1/2}\otimes I_n)$,
with $\widehat{\boldsymbol{\Sigma}}^{-1/2}$ symmetric. The second equality is the definition of the squared Euclidean norm, and the third follows from $\vect(\mathbf{A}\mathbf{X}\mathbf{B}) = (\mathbf{B}^\top\otimes\mathbf{A})\vect(\mathbf{X})$ with $\mathbf{B}=\widehat{\boldsymbol{\Sigma}}^{-1/2}$. For the last equality, note that
$\mathbf{E}\,\widehat{\boldsymbol{\Sigma}}^{-1/2} = \mathbf{C}\boldsymbol{\Psi}^{1/2}\widehat{\boldsymbol{\Sigma}}^{-1/2} - \mathbf{x}\mathbf{b}^\top\boldsymbol{\Psi}^{1/2}\widehat{\boldsymbol{\Sigma}}^{-1/2}$, and applying the vec identity again yields
$\vect(\mathbf{x}\mathbf{b}^\top\boldsymbol{\Psi}^{1/2}\widehat{\boldsymbol{\Sigma}}^{-1/2}) = \big((\boldsymbol{\Psi}^{1/2}\widehat{\boldsymbol{\Sigma}}^{-1/2})^\top\otimes\mathbf{x}\big)\mathbf{b} = (\widehat{\boldsymbol{\Sigma}}^{-1/2}\boldsymbol{\Psi}^{1/2}\otimes\mathbf{x})\mathbf{b}$.
The right-hand side of \eqref{eq:fos-genleastsquares} is the least-squares term of \eqref{eq:frm} with $\mathbf{u}$, $\mathbf{Z}$ and $\boldsymbol{\delta}$ as in \eqref{eq:uZfos-gls}.

\begin{remark}[Estimation of $\boldsymbol{\Sigma}$]\label{rem:whitening}
The matrix $\boldsymbol{\Sigma}$ is estimated from the first-stage residuals in three steps.
\begin{enumerate}
    \item Compute the residual coefficients $\widehat{\boldsymbol{\rho}}_i=\boldsymbol{\Psi}^{1/2}\left(\mathbf{c}_i-x_i\widehat{\mathbf{b}}\right)$, $i=1,\ldots,n$, where $\widehat{\mathbf{b}}$ is the estimate obtained under uncorrelated errors.
    \item Compute their sample covariance matrix
    \[
    \mathbf{S}=\frac{1}{n}\sum_{i=1}^n\widehat{\boldsymbol{\rho}}_i\widehat{\boldsymbol{\rho}}_i^\top=\frac{1}{n}\widehat{\boldsymbol{\varepsilon}}^\top\widehat{\boldsymbol{\varepsilon}},
    \]
    and its eigendecomposition $\mathbf{S}=\mathbf{V}\boldsymbol{\Lambda}\mathbf{V}^\top$.
    \item Set $\widehat{\boldsymbol{\Sigma}}=\mathbf{V}\boldsymbol{\Lambda}\mathbf{V}^\top$ and $\widehat{\boldsymbol{\Sigma}}^{-1/2}=\mathbf{V}\boldsymbol{\Lambda}^{-1/2}\mathbf{V}^\top$.
\end{enumerate}
The factor $\boldsymbol{\Psi}^{1/2}$ in $\widehat{\boldsymbol{\rho}}_i$ is what makes this eigenanalysis the FPCA of the residual curves: because the basis functions in $\mathbf{B}(t)$ are not orthonormal, the Euclidean norm of the raw coefficients $\mathbf{c}_i-x_i\widehat{\mathbf{b}}$ differs from the $L^2$ norm of the curve, whereas $\|\widehat{\boldsymbol{\rho}}_i\|_2$ equals it. Hence the eigenvalues of $\widehat{\boldsymbol{\Sigma}}$ are the variances of the functional principal components, and the eigenvectors give their coefficients in the orthonormal basis $\mathbf{B}(t)\boldsymbol{\Psi}^{-1/2}$.
\end{remark}

\section{Estimation of the general FOF linear regression model~\texorpdfstring{\eqref{eq:fof}}{}
         \label{app:FOF-general}}

In this appendix we detail the calculations leading to the expressions of $\mathbf{u}$ and
$\mathbf{Z}$ in \eqref{eq:uZfof} and \eqref{eq:uZfof-gls} for the FOF linear regression model \eqref{eq:fof},
when the basis representations $x_i(s)=\boldsymbol{\Gamma}(s)\mathbf{a}_i$,
$y_i(t)=\mathbf{B}(t)\mathbf{c}_i$ and
$\beta(s,t)=\mathbf{B}(t)\mathcal{B}\boldsymbol{\Gamma}(s)^\top$ in \eqref{eq:xiB-spline},
\eqref{eq:yiB-spline} and \eqref{eq:fofbetaB-spline} are assumed.

\subsection{Fitted term and residual}

Substituting the basis representations of the covariate and of the coefficient surface into
\eqref{eq:fof}, and using that $\mathcal{B}$ does not depend on $s$,
\begin{equation}\label{eq:fitted-fof}
\int_0^T x_i(s)\beta(s,t)\,ds
= \mathbf{B}(t)\mathcal{B}\int_0^T \boldsymbol{\Gamma}(s)^\top \boldsymbol{\Gamma}(s)\,ds\;\mathbf{a}_i
= \mathbf{B}(t)\mathcal{B}\,\boldsymbol{\Psi}_{\Gamma}\mathbf{a}_i
= \mathbf{B}(t)\mathcal{B}\,\mathbf{w}_i,
\end{equation}
with $\mathbf{w}_i=\boldsymbol{\Psi}_{\Gamma}\mathbf{a}_i$ as in \eqref{eq:wi}. Both the fitted term
\eqref{eq:fitted-fof} and the response $y_i(t)=\mathbf{B}(t)\mathbf{c}_i$ belong to the linear span of
$\mathbf{B}(t)$, so the residual function of \eqref{eq:fof-basis} is simply
\begin{equation}\label{eq:residual-fof}
r_i(t) \;=\; \mathbf{B}(t)\left(\mathbf{c}_i-\mathcal{B}\mathbf{w}_i\right),
\qquad i=1,\ldots,n,
\end{equation}
and no change of representation is needed, in contrast with the concurrent model of
Appendix~\ref{app:FOF-concurrent}.

\subsection{The IMSE under uncorrelated errors}

Let $\mathbf{C},\mathbf{W}\in\mathbb{R}^{n\times(k+d)}$ be the matrices whose $i$-th rows are
$\mathbf{c}_i^\top$ and $\mathbf{w}_i^\top$, respectively, so that the $i$-th row of
$\mathbf{C}-\mathbf{W}\mathcal{B}^\top$ is $\left(\mathbf{c}_i-\mathcal{B}\mathbf{w}_i\right)^\top$.
Proceeding as in Appendix~\ref{app:FOS-errorsdep}, and denoting by $\|\cdot\|_F$ the Frobenius norm,
\begin{align}
\sum_{i=1}^n \int_0^T r_i(t)^2\,dt
&= \int_0^T \left\| \mathbf{B}(t)\left(\mathbf{C}-\mathbf{W}\mathcal{B}^\top\right)^\top
\right\|_2^2 dt \nonumber\\
&= \int_0^T \Tr\left\{ \mathbf{B}(t)\left(\mathbf{C}-\mathbf{W}\mathcal{B}^\top\right)^\top
\left(\mathbf{C}-\mathbf{W}\mathcal{B}^\top\right)\mathbf{B}(t)^\top \right\} dt \nonumber\\
&= \Tr\left\{ \left(\mathbf{C}-\mathbf{W}\mathcal{B}^\top\right)^\top
\left(\mathbf{C}-\mathbf{W}\mathcal{B}^\top\right)
\int_0^T\mathbf{B}(t)^\top\mathbf{B}(t)\,dt \right\} \nonumber\\
&= \Tr\left\{ \left(\mathbf{C}-\mathbf{W}\mathcal{B}^\top\right)^\top
\left(\mathbf{C}-\mathbf{W}\mathcal{B}^\top\right)\boldsymbol{\Psi} \right\}
\;=\; \left\|\left(\mathbf{C}-\mathbf{W}\mathcal{B}^\top\right)\boldsymbol{\Psi}^{1/2}\right\|_F^2
\nonumber\\
&= \left\| \vect\left(\mathbf{C}\boldsymbol{\Psi}^{1/2}\right)
- \vect\left(\mathbf{W}\mathcal{B}^\top\boldsymbol{\Psi}^{1/2}\right) \right\|_2^2
\nonumber\\
&= \left\| \vect\left(\mathbf{C}\boldsymbol{\Psi}^{1/2}\right)
- \left(\boldsymbol{\Psi}^{1/2}\otimes\mathbf{W}\right)\vect\left(\mathcal{B}^\top\right)
\right\|_2^2, \label{eq:imse-fof-app}
\end{align}
where the second equality uses the invariance of the trace under cyclic permutations, and the last one
follows from $\vect(\mathbf{A}\mathbf{X}\mathbf{B})=(\mathbf{B}^\top\otimes\mathbf{A})\vect(\mathbf{X})$
with $\mathbf{A}=\mathbf{W}$, $\mathbf{X}=\mathcal{B}^\top$ and
$\mathbf{B}=\boldsymbol{\Psi}^{1/2}=(\boldsymbol{\Psi}^{1/2})^\top$. Expression
\eqref{eq:imse-fof-app} is the least squares term $\|\mathbf{u}-\mathbf{Z}\boldsymbol{\delta}\|_2^2$ of
\eqref{eq:frm} with $\mathbf{u}$, $\mathbf{Z}$ and $\boldsymbol{\delta}$ as in \eqref{eq:uZfof}. Note
that \eqref{eq:imse-fof-app} coincides with \eqref{eq:fos_imse_indep} after replacing the vector
$\mathbf{x}$ of scalar covariates by the matrix $\mathbf{W}$ and $\mathbf{b}$ by
$\vect(\mathcal{B}^\top)$.

\subsection{The GLS criterion under correlated errors}

Let $\mathbf{E}=\left(\mathbf{C}-\mathbf{W}\mathcal{B}^\top\right)\boldsymbol{\Psi}^{1/2}
\in\mathbb{R}^{n\times(k+d)}$, whose $i$-th row is $\boldsymbol{\rho}_i^\top$ with
$\boldsymbol{\rho}_i$ as in \eqref{eq:rho} but with $x_i\mathbf{b}$ replaced by
$\mathcal{B}\mathbf{w}_i$. Since the residual curves are independent across individuals with common
covariance matrix $\boldsymbol{\Sigma}=\mathrm{Cov}(\boldsymbol{\rho}_i)$, it holds that
$\mathrm{Cov}\left(\vect(\mathbf{E})\right)=\boldsymbol{\Sigma}\otimes I_n$, exactly as in the FOS
case. Estimating $\boldsymbol{\Sigma}$ by FPCA on the residual curves of the first stage
\citep{yao2005functional}, as described in Remark~\ref{rem:whitening}, the GLS criterion reads
\begin{align}
& \left(\vect\left(\mathbf{C}\boldsymbol{\Psi}^{1/2}\right)
- \left(\boldsymbol{\Psi}^{1/2}\otimes\mathbf{W}\right)\boldsymbol{\delta}\right)^\top
\left(\widehat{\boldsymbol{\Sigma}}\otimes I_n\right)^{-1}
\left(\vect\left(\mathbf{C}\boldsymbol{\Psi}^{1/2}\right)
- \left(\boldsymbol{\Psi}^{1/2}\otimes\mathbf{W}\right)\boldsymbol{\delta}\right) \nonumber\\
&\qquad = \left\|\vect\left(\mathbf{E}\,\widehat{\boldsymbol{\Sigma}}^{-1/2}\right)\right\|_2^2
\;=\; \left\| \vect\left(\mathbf{C}\boldsymbol{\Psi}^{1/2}
\widehat{\boldsymbol{\Sigma}}^{-1/2}\right)
- \left(\widehat{\boldsymbol{\Sigma}}^{-1/2}\boldsymbol{\Psi}^{1/2}\otimes\mathbf{W}\right)
\boldsymbol{\delta}\right\|_2^2, \label{eq:fof-genleastsquares}
\end{align}
which is the least squares term of \eqref{eq:frm} with $\mathbf{u}$, $\mathbf{Z}$ and
$\boldsymbol{\delta}$ as in \eqref{eq:uZfof-gls}. The steps are those of
\eqref{eq:fos-genleastsquares} with $\mathbf{x}$ replaced by $\mathbf{W}$.

\subsection{The solution}

Adding the penalty \eqref{eq:penalty2d} turns \eqref{eq:frm} into an unconstrained convex quadratic
problem whose solution is
\begin{equation}\label{eq:solution-fof}
\widehat{\boldsymbol{\delta}}
= \left(\mathbf{Z}^\top\mathbf{Z}
+ \lambda_t\,\mathbf{D}^\top\mathbf{D}\otimes I_{k+d}
+ \lambda_s\,I_{k+d}\otimes\mathbf{D}^\top\mathbf{D}\right)^{-1}\mathbf{Z}^\top\mathbf{u},
\end{equation}
and the estimated coefficient surface is
$\widehat{\beta}(s,t)=\mathbf{B}(t)\widehat{\mathcal{B}}\boldsymbol{\Gamma}(s)^\top$, where
$\vect(\widehat{\mathcal{B}}^\top)=\widehat{\boldsymbol{\delta}}$.

\begin{remark}[Kronecker structure of the least squares term]\label{rem:kronecker}
Under uncorrelated errors, the matrices $\mathbf{Z}^\top\mathbf{Z}$ and $\mathbf{Z}^\top\mathbf{u}$
that determine the solution \eqref{eq:solution-fof} inherit a Kronecker structure. Indeed, the
mixed-product rule
$(\mathbf{A}\otimes\mathbf{B})(\mathbf{C}\otimes\mathbf{D})=\mathbf{A}\mathbf{C}\otimes\mathbf{B}\mathbf{D}$
gives
\begin{equation}\label{eq:kronfof}
\mathbf{Z}^\top\mathbf{Z}
= \left(\boldsymbol{\Psi}^{1/2}\otimes\mathbf{W}\right)^\top
\left(\boldsymbol{\Psi}^{1/2}\otimes\mathbf{W}\right)
= \boldsymbol{\Psi}\otimes\mathbf{W}^\top\mathbf{W},
\qquad
\mathbf{Z}^\top\mathbf{u}
= \vect\left(\mathbf{W}^\top\mathbf{C}\,\boldsymbol{\Psi}\right),
\end{equation}
so that neither $\mathbf{Z}$ nor $\mathbf{Z}^\top\mathbf{Z}$ needs to be formed explicitly, and the
array algorithms of \citet{currie2006generalized} can be exploited when solving
\eqref{eq:solution-fof}. Moreover, since
$\mathrm{rank}\left(\mathbf{Z}^\top\mathbf{Z}\right)=(k+d)\,\mathrm{rank}(\mathbf{W})
\leq (k+d)\min\{n,k+d\}$, the unpenalized problem is rank deficient whenever $n<k+d$; in that regime
the penalty \eqref{eq:penalty2d} is not only a smoothing device but also what makes the estimator
well defined.
\end{remark}

\section{Estimation of the concurrent FOF linear regression model~\texorpdfstring{\eqref{eq:fofconcurrent}}{} \label{app:FOF-concurrent}}

In this appendix we detail the calculations leading to the expressions of $\mathbf{u}$ and
$\mathbf{Z}$ in \eqref{eq:uZconcurrent} for the concurrent FOF linear regression model
\eqref{eq:fofconcurrent}, when the basis representations $x_i(t)=\mathbf{B}(t)\mathbf{a}_i$,
$y_i(t)=\mathbf{B}(t)\mathbf{c}_i$ and $\beta(t)=\mathbf{B}(t)\mathbf{b}$ in \eqref{eq:xiB-spline},
\eqref{eq:yiB-spline} and \eqref{eq:sofbetaB-spline} are assumed. Note that here the covariate and the
response are necessarily expanded in the same basis, since $s=t$.

\subsection{A common tensor-product representation}

The fitted term $x_i(t)\beta(t)$ in \eqref{eq:fofconcurrent} is a product of two splines of degree
$d$ and, therefore, a piecewise polynomial of degree $2d$ that does not belong to the linear span of
$\mathbf{B}(t)$, as noted in Section~\ref{subsubsec:concurrent}. Consequently, the coefficients of
the response and those of the fitted term cannot be matched in the original basis, and both are
expressed in the common tensor-product family
$\{B_j(t)B_\ell(t)\}_{1\leq j,\ell\leq k+d}$, whose row vector is the Kronecker product
$\mathbf{B}(t)\otimes\mathbf{B}(t)\in\mathbb{R}^{1\times(k+d)^2}$.

\paragraph{Fitted term.}
Since $\mathbf{B}(t)\mathbf{a}_i$ is a scalar, and using the identities
$\vect(\mathbf{A}\mathbf{X}\mathbf{B})=(\mathbf{B}^{\top}\!\otimes \mathbf{A})\vect(\mathbf{X})$ and
$\vect(\mathbf{u}\mathbf{v}^{\top})=\mathbf{v}\otimes\mathbf{u}$, the fitted term in
\eqref{eq:fofconcurrent} can be written as
\begin{eqnarray}
x_i(t)\beta(t)
&=& \left(\mathbf{B}(t)\mathbf{b}\right)\left(\mathbf{B}(t)\mathbf{a}_i\right)
= \mathbf{B}(t)\,\mathbf{b}\mathbf{a}_i^{\top}\mathbf{B}(t)^{\top} \nonumber\\
&=& \left(\mathbf{B}(t)\otimes\mathbf{B}(t)\right)(\mathbf{a}_i\otimes\mathbf{b})
= \left(\mathbf{B}(t)\otimes\mathbf{B}(t)\right)(\mathbf{a}_i\otimes I_{k+d})\,\mathbf{b},
\label{eq:fitted-concurrent}
\end{eqnarray}
where the last equality follows from
$\mathbf{a}_i\otimes\mathbf{b}=(\mathbf{a}_i\otimes I_{k+d})\mathbf{b}$.

\paragraph{Response.}
The response is lifted to the same family of functions by exploiting the partition-of-unity property
of B-spline bases, namely
\begin{equation*}
\mathbf{B}(t)\,\mathbf{1} = \sum_{j=1}^{k+d} B_j(t) = 1
\qquad \text{for all } t\in[0,T],
\end{equation*}
where $\mathbf{1}\in\mathbb{R}^{k+d}$ is the vector of ones. Then,
\begin{equation}\label{eq:response-concurrent}
y_i(t)
= \left(\mathbf{B}(t)\mathbf{c}_i\right)\left(\mathbf{B}(t)\mathbf{1}\right)
= \mathbf{B}(t)\,\mathbf{c}_i\mathbf{1}^{\top}\mathbf{B}(t)^{\top}
= \left(\mathbf{B}(t)\otimes\mathbf{B}(t)\right)(\mathbf{1}\otimes\mathbf{c}_i).
\end{equation}

\paragraph{Residual.}
Combining \eqref{eq:fitted-concurrent} and \eqref{eq:response-concurrent}, the residual function of
model \eqref{eq:fofconcurrent} admits the compact expression
\begin{equation}\label{eq:residual-concurrent}
r_i(t) \;=\; y_i(t)-x_i(t)\beta(t)
= \left(\mathbf{B}(t)\otimes\mathbf{B}(t)\right)\mathbf{v}_i,
\qquad
\mathbf{v}_i = \mathbf{1}\otimes\mathbf{c}_i-(\mathbf{a}_i\otimes I_{k+d})\,\mathbf{b},
\end{equation}
for $i=1,\ldots,n$.

\subsection{The IMSE under uncorrelated errors}

As in the FOS and general FOF cases, the least squares criterion corresponds to minimizing an IMSE
over the domain $[0,T]$. Using \eqref{eq:residual-concurrent},
\begin{equation}\label{eq:imse-concurrent}
\begin{aligned}
\sum_{i=1}^{n}\int_0^T r_i(t)^2\, dt
&= \sum_{i=1}^{n}\mathbf{v}_i^\top
\left(\int_0^T\left(\mathbf{B}(t)\otimes\mathbf{B}(t)\right)^\top
\left(\mathbf{B}(t)\otimes\mathbf{B}(t)\right)dt\right)\mathbf{v}_i
\;=\; \sum_{i=1}^{n}\mathbf{v}_i^{\top}\bar{\boldsymbol{\Psi}}\,\mathbf{v}_i\\[3pt]
&= \sum_{i=1}^{n}
\left\|\,\bar{\boldsymbol{\Psi}}^{1/2}\left(\mathbf{1}\otimes\mathbf{c}_i\right)
-\bar{\boldsymbol{\Psi}}^{1/2}\left(\mathbf{a}_i\otimes I_{k+d}\right)\mathbf{b}\,\right\|_2^2,
\end{aligned}
\end{equation}
where
\begin{equation}\label{eq:psibar}
\bar{\boldsymbol{\Psi}}
= \int_0^T \left(\mathbf{B}(t)\otimes\mathbf{B}(t)\right)^{\top}
\left(\mathbf{B}(t)\otimes\mathbf{B}(t)\right)dt
\;\in\;\mathbb{R}^{(k+d)^2\times(k+d)^2},
\end{equation}
with entries of the form $\int_0^T B_j(t)B_\ell(t)B_m(t)B_r(t)\,dt$, is the fourth-order analogue of
the inner product matrix $\boldsymbol{\Psi}$. It is symmetric and positive semidefinite and,
therefore, admits a symmetric square root $\bar{\boldsymbol{\Psi}}^{1/2}$. Stacking \eqref{eq:imse-concurrent} over the $n$
observations yields the least squares term $\|\mathbf{u} - \mathbf{Z}\boldsymbol{\delta}\|_2^2$ in
\eqref{eq:frm}, with $\mathbf{u}$, $\mathbf{Z}$ and $\boldsymbol{\delta} = \mathbf{b}$ as given in
\eqref{eq:uZconcurrent}. Adding the penalty term
$P(\boldsymbol{\delta})=\lambda\|\mathbf{D}\boldsymbol{\delta}\|_2^2$ turns \eqref{eq:frm} into an
unconstrained convex quadratic problem whose solution is
\begin{equation}\label{eq:solution-concurrent}
\widehat{\mathbf{b}}
= \left(\mathbf{Z}^{\top}\mathbf{Z} + \lambda \mathbf{D}^{\top}\mathbf{D}\right)^{-1}
\mathbf{Z}^{\top}\mathbf{u},
\end{equation}
and the estimated coefficient function is $\widehat{\beta}(t)=\mathbf{B}(t)\widehat{\mathbf{b}}$.

\subsection{The GLS criterion under correlated errors}

Set $\boldsymbol{\rho}_i=\bar{\boldsymbol{\Psi}}^{1/2}\mathbf{v}_i$, with $\mathbf{v}_i$ as in
\eqref{eq:residual-concurrent}, so that the $i$-th block of the residual vector associated with
\eqref{eq:uZconcurrent} is $\boldsymbol{\rho}_i$ and, by \eqref{eq:imse-concurrent},
$\int_0^T r_i(t)^2\,dt=\|\boldsymbol{\rho}_i\|_2^2$. Because those blocks are stacked by individual,
and the residual curves are independent across individuals, the covariance matrix of the stacked
residual vector is $I_n\otimes\boldsymbol{\Sigma}$ with
$\boldsymbol{\Sigma}=\mathrm{Cov}(\boldsymbol{\rho}_i)$, and the GLS criterion is
\begin{equation}\label{eq:concurrent-gls}
\sum_{i=1}^n \left\| \widehat{\boldsymbol{\Sigma}}^{-1/2}\bar{\boldsymbol{\Psi}}^{1/2}
\left(\mathbf{1}\otimes\mathbf{c}_i\right)
- \widehat{\boldsymbol{\Sigma}}^{-1/2}\bar{\boldsymbol{\Psi}}^{1/2}
\left(\mathbf{a}_i\otimes I_{k+d}\right)\mathbf{b} \right\|_2^2,
\end{equation}
that is, \eqref{eq:frm} with the blocks of \eqref{eq:uZconcurrent} pre-multiplied by
$\widehat{\boldsymbol{\Sigma}}^{-1/2}$, and with $\widehat{\boldsymbol{\Sigma}}$ estimated by FPCA on
the residual curves of the first stage \citep{yao2005functional}. Note also that the
Kronecker factors appear in reverse order with respect to the FOS and general FOF cases, where the
residuals are stacked by basis index rather than by individual; this is immaterial, as both stackings
are related by a permutation.

\begin{remark}[Computational aspects and the raw-data formulation]\label{rem:normal-concurrent}
The mixed-product rule
$(\mathbf{A}\otimes\mathbf{B})(\mathbf{C}\otimes\mathbf{D})=\mathbf{A}\mathbf{C}\otimes\mathbf{B}\mathbf{D}$
gives
\begin{equation*}
\begin{aligned}
&\left(\mathbf{B}(t)\otimes\mathbf{B}(t)\right)\left(\mathbf{a}_i\otimes I_{k+d}\right)
= \left(\mathbf{B}(t)\mathbf{a}_i\right)\otimes\mathbf{B}(t) = x_i(t)\,\mathbf{B}(t),\\[1ex]
&\left(\mathbf{B}(t)\otimes\mathbf{B}(t)\right)\left(\mathbf{1}\otimes\mathbf{c}_i\right)
= y_i(t),
\end{aligned}
\end{equation*}
so that, with $\mathbf{u}$ and $\mathbf{Z}$ as in \eqref{eq:uZconcurrent},
\begin{equation*}
\mathbf{Z}^\top\mathbf{Z}
= \sum_{i=1}^n\left(\mathbf{a}_i\otimes I_{k+d}\right)^\top\bar{\boldsymbol{\Psi}}
\left(\mathbf{a}_i\otimes I_{k+d}\right)
= \sum_{i=1}^{n}\int_0^T x_i(t)^2\,\mathbf{B}(t)^{\top}\mathbf{B}(t)\,dt,
\end{equation*}
\begin{equation*}
\mathbf{Z}^\top\mathbf{u}
= \sum_{i=1}^n\left(\mathbf{a}_i\otimes I_{k+d}\right)^\top\bar{\boldsymbol{\Psi}}
\left(\mathbf{1}\otimes\mathbf{c}_i\right)
= \sum_{i=1}^{n}\int_0^T x_i(t)\,y_i(t)\,\mathbf{B}(t)^{\top}\,dt,
\end{equation*}
which are the expressions given in \eqref{eq:normalconcurrent}. They coincide with those obtained by
differentiating the IMSE \eqref{eq:imse-concurrent} directly with respect to $\mathbf{b}$, and
establish the connection with the raw-data formulation discussed in Remark~\ref{rem:rawdata}.
\end{remark}

\begin{remark}[Evaluation of the tensor-product family]\label{rem:rowkron}
The tensor-product family $\{B_j(t)B_\ell(t)\}_{1\leq j,\ell\leq k+d}$ is introduced in the continuous
setting through the row vector $\mathbf{B}(t)\otimes\mathbf{B}(t)\in\mathbb{R}^{1\times (k+d)^2}$.
After evaluating the basis functions at the observation points $t_1,\ldots,t_M$, the resulting design
matrix is
\[
\widetilde{\mathbf{B}}_{\otimes} =
\begin{pmatrix}
	\mathbf{B}(t_1)\otimes\mathbf{B}(t_1)\\
	\vdots\\
	\mathbf{B}(t_M)\otimes\mathbf{B}(t_M)
\end{pmatrix}
\in\mathbb{R}^{M\times (k+d)^2},
\]
whose $m$-th row contains all pairwise products $B_j(t_m)B_\ell(t_m)$. This matrix should not be
confused with the Kronecker product of the evaluation matrix with itself,
$\mathbf{B}\otimes\mathbf{B}\in\mathbb{R}^{M^2\times (k+d)^2}$, which contains all combinations of
pairs of observation points $(t_r,t_s)$. Rather, it corresponds to the row-wise evaluation of the
tensor-product family at the common points $t_1,\ldots,t_M$. In matrix computation, this object is
often referred to as the row-wise Kronecker product (or row tensor product) of the evaluation matrix
with itself.
\end{remark}

\begin{remark}[Numerical evaluation of $\bar{\boldsymbol{\Psi}}$]\label{rem:psibar-quadrature}
The matrix $\bar{\boldsymbol{\Psi}}$ involves integrals of products of four $B$-spline basis functions.
In practice, it can be approximated by numerical quadrature. Let $t_1,\ldots,t_M$ be an equally spaced
grid on $[0,T]$, with $M$ odd, and let
$\widetilde{\mathbf{B}}_{\otimes}\in\mathbb{R}^{M\times (k+d)^2}$ be the row-wise Kronecker product
defined in Remark~\ref{rem:rowkron}. Using the composite Simpson rule, with weights
\[
\boldsymbol{\omega}=\frac{h}{3}(1,4,2,4,\ldots,2,4,1),
\qquad h=\frac{T}{M-1},
\]
we approximate
$\bar{\boldsymbol{\Psi}}\approx\widetilde{\mathbf{B}}_{\otimes}^{\top}\boldsymbol{\Omega}
\widetilde{\mathbf{B}}_{\otimes}$ with $\boldsymbol{\Omega}=\mathrm{diag}(\omega_1,\ldots,\omega_M)$. Equivalently, each
entry of $\bar{\boldsymbol{\Psi}}$ is approximated by applying Simpson's rule to the corresponding
product $B_j(t)B_\ell(t)B_r(t)B_s(t)$. 
\end{remark}

\end{document}